\documentclass[11pt]{article}
\makeatletter
\usepackage{authblk}
\makeatother

\usepackage[utf8]{inputenc}
\usepackage[english]{babel}

\usepackage{comment}
\usepackage{natbib}
\usepackage{bm,lscape}
\usepackage{lscape}
\usepackage{mathrsfs,dsfont}
\usepackage{longtable}
\usepackage{graphicx}
\usepackage[onehalfspacing]{setspace}

\usepackage{amsmath,amsthm}
\usepackage{amssymb}
\usepackage{multirow}
\usepackage{bbm}
\usepackage{bm}
\usepackage{hyperref}
\usepackage{cleveref}
\usepackage{mathrsfs}
\usepackage{dsfont}
\usepackage{color}
\usepackage{subfig}
\usepackage{tikz}
\usetikzlibrary{arrows}
\usetikzlibrary{automata,positioning}

\usepackage{caption}
\allowdisplaybreaks  
\title{The survival of start-ups in time of crisis. A machine learning approach to measure innovation}

\author[1, 5]{Marco Guerzoni}
\author[2, 1, 3]{Consuelo R. Nava}
\author[4, 1]{Massimiliano Nuccio}

\affil[1]{Despina, Department of Economics and Statistics, University of Turin}
\affil[2]{Department of Economics and Political Sciences, University of Aosta Valley}
\affil[3]{Department of Economic Policy, Universit\`a Cattolica del Sacro Cuore di Milano}
\affil[4]{City REDI - University of Birmingham\footnote{Corresponding author: m.nuccio@bham.ac.uk}}
\affil[5]{ICRIOS, Bocconi University}

\begin{document}

\maketitle

\begin{abstract}
\noindent 
This paper shows how data science can contribute to improving empirical research in economics by leveraging on large datasets and extracting information otherwise unsuitable for a traditional econometric approach. As a test-bed for our framework, machine learning algorithms allow us to create a new holistic measure of innovation built on a 2012 Italian Law aimed at boosting new high-tech firms. We adopt this measure to analyse the impact of innovativeness on a large population of Italian firms which entered the market at the beginning of the 2008 global crisis. 
The methodological contribution is organised in different steps. First, we train seven supervised learning algorithms to recognise innovative firms on 2013 firmographics data and select a combination of those with best predicting power. Second, we apply the former on the 2008 dataset and predict which firms would have been labelled as innovative according to the definition of the law. Finally, we adopt this new indicator as regressor in a survival model to explain firms’ ability to remain in the market after 2008. Results suggest that the group of innovative firms are more likely to survive than the rest of the sample, but the survival premium is likely to depend on location.

\vspace{0.5cm}
JEL CODES: O30, D22, C52, C55.


\end{abstract}

\thispagestyle{empty}
\newpage

\section{Introduction}{\label{sec:intro}}

This paper shows how data science can contribute to empirical research in economics by leveraging on large datasets and extracting information otherwise unsuitable for a traditional econometric approach. Yet, research questions drawn from the economic theory, on the one hand, and assumptions in econometric modelling, on the other, guide our choices to exploit the richness of the science of data. As exemplary case and further contribution to the literature, we apply this framework to evaluate performances and survival rate of innovative start-ups (hereafter, INNs) vis-\`a-vis other types of newly funded firms (non-innovative start-ups, hereafter NOINNs) for which empirical evidences show controversial results. More consensus can be found around the two major challenges which undermine a robust causal relationship between innovation and survival probability.  First, most commonly selected proxies and measures for innovation have revealed serious limitations in capturing innovation \citep{manual}. Second, firm survival may depend on many internal and external factors, therefore the innovation effect is not easy to isolate and might suffer from confounding issues \citep{freeman1994}.
Nevertheless, this paper does not want to be just another study of the innovation effect on firm survival. Our contribution, indeed, is primarily methodological. We adopt an alternative and holistic measure of innovation drawn from the Italian national regulation. Therefore, we analyse the effect of innovation on the survival probability of a large sample of Italian start-ups established in 2008, the very first year of the financial crisis that marked a strong acceleration of the Italian industrial decline. Assuming that the crisis exacerbated both market risks and financial constraints, this database offers an extraordinary opportunity of testing the effect of a very strong selection mechanism. If there is any truth in the evolutionary framework, which describes industrial dynamics as triggered by the evolutionary mechanism of entry and selection, we should be able to observe it in a time of crisis.

Our empirical strategy is able to effectively relax some of the constraints imposed by the traditional inferential analysis by integrating a data science approach with econometrics, according to the following three steps.

First, we adopt a definition of ``innovative start-up'' built on the multiple criteria prescribed by the Italian regulation in 2012 aimed at boosting new high-tech firms through a program of incentives. Therefore, we extract all available new entrant firms in 2013 from AIDA, the Bureau Van Dijk database, including start-ups both registered and not registered as innovative according to the above regulation. After a data cleansing process, we implement a supervised machine learning approach based on the training of seven algorithms (namely classification and regression trees, logistic regression, na\"ive Bayesian classifier and artificial neural network) to predict the probability of being  INNs using 124 firmographics variables. Since the innovation literature considers sectors and locations important confounding effects in explaining survival, we exclude them from the training-set of the machine learning algorithms. This allows us to eventually include these variables in an econometric framework, without the risk of describing spurious relationships.

Second, from the same database, we extract the sample of new firms entering the market in 2008, which faced the highly selective environment of the crisis, and we select a combination of the above algorithms able to predict  the probability of being INNs. 

Third, once we can discriminate between INNs and NOINNs according to the above multi-criteria definition, we estimate with a Cox proportional hazards model  firms' survival over ten years (2008-2018), controlling for the impact of sectors and locations. 
Without the use of machine learning algorithms, this innovative measure of innovation could have not been created and, without a clear theoretical input from the literature and the econometric assumptions to guide the machine learning modelling, this new indicator would have been useless. 

The paper proceeds as follows. 
In Session~\ref{sec:bigdata}, we present our methodological approach and explain how it can contribute to economic empirical analyses in general, while  Session~\ref{sec:inno} positions our contribution in the debate around the role of innovation in fostering survival in the market, as a specific case to test our methodology.
Thereafter, in Section~\ref{data}, we present the machine learning process which leads to the creation of the new indicator for innovation. In Section~\ref{analysis}, we carry out a survival analysis while in Section~\ref{sec:conclu} we summarise and discuss the main results of the paper as well as the new challenges ahead.


\section{Data science: an opportunity for the creation of new variables}\label{sec:bigdata}

The data science paradigm consists of the convergence of complementary technologies which, when combined, allow the extrapolation of information and knowledge from very large dataset: algorithms, computational power, collection and storage of digitised data \citep{estolatan2018mapping}.
Along with \citet{Varian}, this paradigmatic change has provided economists with an expanded set of analytical tools to explore data and acquire information. In particular, we can recognise at least three types of approaches to data analysis which widely differ among each other both in the goals and in the way they test the uncertainty of a model. 

Econometric analysis is the oldest and most popular one, also for not strictly economic problems, and it aims at highlighting causal relations between variables. The external validity of its results relies on statistical inference, which requires available observations to be a random sample of the population. If not, well-known techniques have been developed for non-random data or for the generation of truly random data in experimental settings. If the assumptions for the statistical inference are fulfilled, the researcher can suitably estimate the average effect as well as model the associated uncertainty of the phenomenon of interest in the population. However, this result comes with a cost attached. Estimator properties, which allow for a suitable statistical inference, have been derived on a limited class of mostly linear models. Moreover, their statistical derivation imposes limits on the ability to model the complexity of the phenomenon of interest. Feedback between variables are difficult to handle and even prohibited between dependent and independent variables; the presence of heteroskedasticity and autocorrelation in the error terms of the econometric model need to be carefully addressed; and an excess of multicollinearity between covariates raises serious inferential issues. All in all, the capability of highlighting statistically robust causal relations heavily constrains the variety of models that can be implemented and this limitation impinges upon the explanatory power and the performance in out-of-sample predictions. Moreover, the complex reality represented by big data rarely fits into the required econometric assumptions, nor the data collection  always happen in controlled settings.  For this reason, econometrics lacks the capability of fully exploiting the information in big data. We claim, here, that a careful integration of machine learning algorithms can improve the exploitability of information and, under certain conditions, make it synthetically available for an econometric modelling.

A first rapidly growing approach in data science is based on machine learning techniques for prediction and/or classification, also known as supervised machine learning \citep[see, among others,][]{kotsiantis2007supervised}. Predictive models learn from historical data and make predictions on new data where we do not know the answer. Technically, predictive modeling is the problem of approximating a mapping function (f) given input data (X) to predict an output value ($y$).In this framework, algorithms are trained on large number of cases and variables (training-set) and learn from a target category to assign new observations. External validity, \emph{i.e.} the variance of the estimates in out-of-sample predictions, is tested on a partition of the available data (test-set), which is hold up and not employed in the  in the learning process, namely for the algorithm prediction over an unobserved category. For this reason and contrary to the econometric approach, any algorithm employed in machine learning is not restricted by any assumption and the only objective function is to maximise the prediction power on the test-set. In this way, the explanatory power of the algorithm can be very high, since no limits to its functional form are imposed, but nothing can be said on the true impact of the single variable on the target one. 
A clear trade-off emerges between the adoption of models aimed at finding causal relations between variables and models aimed at predicting or classifying a phenomenon \citep{Shmueli2010}. The former are cautious in the data selection and needs to be relatively simple in the functional form to approximate data points and to minimise the mean square error of estimators to confirm the underlying theory. Extremely simple models tend not to fit data well enough (under-fitting) and their explanatory power remains limited to the few variables involved, which are not necessarily those explaining the total variability of the phenomenon outside the sample. In other words, they might be unable to account for the complex nature of social phenomena like innovation characterised by the interdependence and interaction of a variety of agents and factors \citep{Antonelli2009}. The latter are meant to gain excellent performances in prediction, but they are blind to spot any causal relation and risk to capture the nopise of data (over-fitting) \footnote{There is a stream of literature which tries to develop models that overcome this trade-off \citep[for instance]{pearl2018book}, but they are more concerned with the creation of artificial general intelligence. For what concerns the statistical learning on data, the trade-off between the prediction error due to simple model (bias in the sense that they could suffer of variable omission or violation of the underling model assumptions) and the variance of estimates in out-of-sample predictions is still binding.}.

The third approach is still based on machine learning, but in the form of unsupervised algorithms which create a partition of the data without any a-priori on the number and type of categories to be generated. Clustering algorithms \citep{macqueen1967some}, self-organizing maps \citep{carlei2014mapping} and, more recently, topic modelling for text analysis of the economic literature \citep{ambrosino2018topic} belong to this group. In this family of algorithms the validation of the model is pursued by an ex-post educated interpretation of the result.

Economic studies can take advantage of the combination of the above mentioned approaches. For economists, the starting point shall always be a theory that has to be tested within a standard econometric framework. Despite someone suggested that the large availability of data, the computational power, and the algorithms decreed the end of the theory towards a pure data-driven type of science \citep{anderson2008end, prensky2009}, other suggested  \citep{kitchin2014,ambrosino2018topic, nuccio2019big, carota2014application, gould1981} that the large availability of data, which reveals the complexity of the relations in the observed reality, calls for more theory. Data and its analysis can still act as a powerful hypotheses-mining engine \citep{jordan1998learning, carota2014application} and provide new theoretical ideas, which nevertheless need to be filtered by a theoretical interpretation effort.

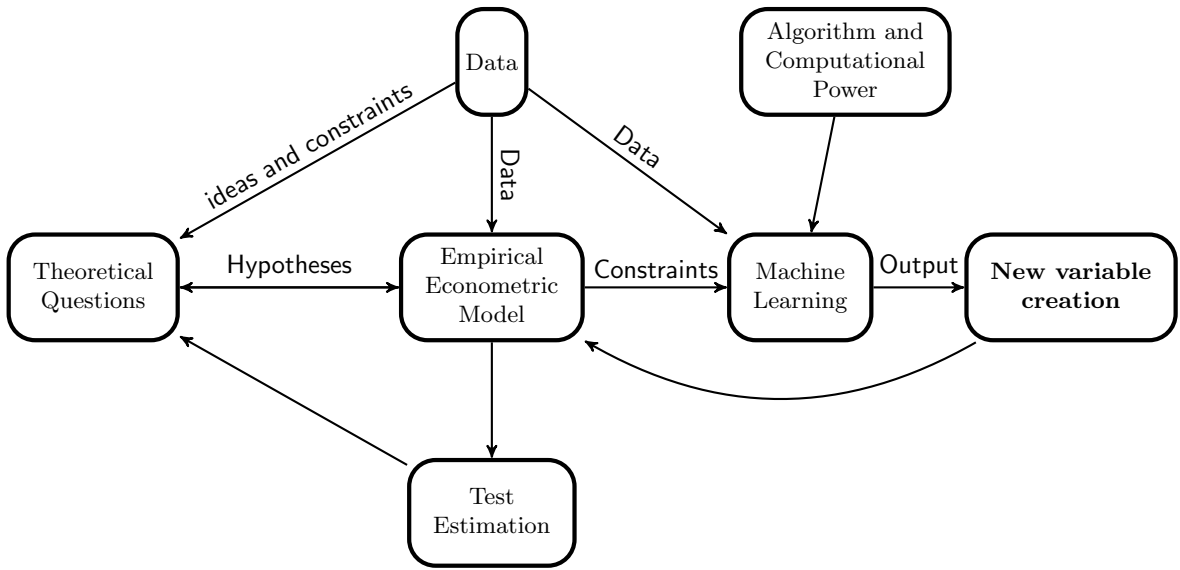
\begin{figure}
    \centering
\begin{footnotesize}
\begin{tikzpicture}[->,>=stealth',auto,node distance=3cm,
  thick,myrectangle/.style={rectangle, draw, minimum height=40, ultra thick, rounded corners=10, black, on grid, auto}
        ]

  \node at (0,0) [myrectangle, draw] (1) {\begin{tabular}{c}Theoretical \\ Questions \end{tabular}};
  \node at (5.3,0)[myrectangle, draw] (2)  {\begin{tabular}{c} Empirical \\ Econometric \\Model\end{tabular}};
  \node at (9.4,0)[myrectangle, draw] (3)  {\begin{tabular}{c}Machine \\ Learning \end{tabular}};
  \node at (13,0)[myrectangle, draw] (4)  {\begin{tabular}{c} \bf{New variable} \\ \bf{creation} \end{tabular}};
  \node at (5.3,3)[myrectangle, draw] (5)  {Data};

\node at (10,3)[myrectangle, draw] (7)  {\begin{tabular}{c}Algorithm and \\ Computational \\ Power\end{tabular}};
  \node at (5.3,-3)[myrectangle, draw] (8)  {\begin{tabular}{c} Test \\ Estimation \end{tabular}};
  \path[every node/.style={font=\sffamily\small}]
    (1) edge node  {Hypotheses} (2)
     (2) edge node [left] {} (1)
    (2) edge node {Constraints} (3)
    (3) edge node {Output} (4)
    (4) edge[bend left] node [left] {} (2)
    (5) edge node [sloped, left, above] {Data} (3)
    (7) edge node [right] {} (3)
      (5) edge node [sloped, above] {Data} (2)
      (5) edge node [sloped, anchor=center, above] {ideas and constraints} (1)
           (2) edge node [right] {} (8)
          (8) edge node [right] {} (1)
      ;

\end{tikzpicture}
\vspace*{5mm}
\end{footnotesize}
\caption{Data science for economics: the creation of a new variable}
    \label{mypicture}
\end{figure}

Within a traditional framework of economic theory and hypotheses testing, the large availability of data can be exploited to create new dependent and independent variables which fit into a standard regression analysis. It should be clear that the theoretical input into the process of data analysis is still a pivotal one, since only the theory can suggest the hypotheses to be tested in a suitable econometric model. Since the latter works properly only under specific assumptions, any econometrics strategy imposes specific properties on the variables to be used, as for instance their types (categorical or continuous), their distribution properties, and their relations with other variables to avoid issues such as multicollinearity or endogeneity in the final analysis. However, once these boundaries are set, data science can employ its brute force, prediction capability, and summarisation potential to leverage on big dataset and extract information that could have not been used in a regression otherwise. In other words, there are both a high complementary and a dense feedback between theory, econometrics and data science. Figure~\ref{mypicture} shows the methodological conceptualisation behind our empirical exercise.

\begin{table}[htbp]\label{tab:recap}
\caption{Summary of the methodology based on econometrics and machine learning tools}
\label{tab:method}
\footnotesize
\begin{tabular}{c|c|c|c|c}
\hline\hline
\textbf{Aim} & \textbf{Tool} & \textbf{Data} & \textbf{Process} & \textbf{Output} \\ \hline
\begin{tabular}[c]{@{}c@{}}Measuring  \\ innovation\end{tabular} & \begin{tabular}[c]{@{}c@{}}Supervised \\machine \\ learning\end{tabular} & \begin{tabular}[c]{@{}c@{}}Data on new Italian \\ firms in 2013 by \\ AIDA Bureau van Dijk\end{tabular} & \begin{tabular}[c]{@{}c@{}}In the data, firms are flagged as \\ innovative if they are registered \\ as INNs according to \\ the 2012 Italian Law 221/2012. \\ The flag is used as a label to \\ train a battery of models  \\  on a subset of the data. \\A second subset is used \\as a test-set to evaluate \\ model performance.\end{tabular} & \begin{tabular}[c]{@{}c@{}}Model trained to \\detect INNs\end{tabular} \\ \hline
\begin{tabular}[c]{@{}c@{}}Controlling for \\ business cycle\end{tabular} & Prediction & \begin{tabular}[c]{@{}c@{}}Data  on new Italian \\ firms in 2008  by \\ AIDA Bureau van Dijk\end{tabular} & \begin{tabular}[c]{@{}c@{}}The supervised algorithm,\\ output of the previous line,\\ is used on 2008 data, \\ during the crisis, \\to detect innovative firms.\end{tabular} & \begin{tabular}[c]{@{}c@{}}Classification of 2008 \\ Italian start-ups between \\ innovative and \\ non-innovative\end{tabular} \\ \hline
\begin{tabular}[c]{@{}c@{}}Controlling for \\sector\\  and\\ location\end{tabular} & \begin{tabular}[c]{@{}c@{}}Multivariate \\ survival \\analysis\end{tabular} & \begin{tabular}[c]{@{}c@{}}Data on a 10 year \\ panel  of new Italian \\ firms in 2008  by \\AIDA Bureau van Dijk, \\ enriched by the \\classification  created \\with the model \end{tabular} & \begin{tabular}[c]{@{}c@{}}We run Cox regressions using \\the classification  of the   \\previous output as main \\independent variable. \\Sector, location and \\ interaction terms \\ are added as controls.\end{tabular} & \begin{tabular}[c]{@{}c@{}}Estimation of the effect \\ of being innovative, \\ sector and location\\  on the survival  \\ probability\end{tabular}
\\
\hline
\hline
\end{tabular}
\end{table}

The test-bed of our approach is rooted in a long-standing controversial evidence on the different survival rates between INNs and NOINNs and on the extent to which this relationship is distorted by a failure in controlling for possible confounding variables. As possible weaknesses in previous works, we highlight  both the type of indicator used to proxy innovation and the lack of consistent controls for sector and location. Following the broad literature on this topic, we further argue that the existence of a survival premium of INNs can be at best tested during the 2008 crisis, when market selection mechanisms were more effective. Only as a second step we turn to data and data science. We thus collect data about new firms in 2013 when a new Italian Law enacted on 17 December 2012 provides incentives to start-ups to be identified as innovative firms. We employ a supervised machine learning approach to estimate the probability of firms in 2013 to belong to the given class of ``innovative start-ups'' and then we apply the same algorithm to predict which firms in 2008 could have been labelled as innovative according to the 2012 law. As explained before, we partition the 2013 start-up sample in the training- and the test-set. On the training, we apply a series of algorithms (see Appendix~\ref{algo}) with different degrees of complexity and with the aim of maximising their prediction power on the test-set. The algorithm, or the combination of algorithms, with the best performance is then used to predict INNs on the 2008 sample. However, to include the new measure of innovation in an econometric model, the predictive algorithm has been trained on all available variables but sector and location, which will be used as covariates. The remaining of the paper discusses the details of this process, which is also briefly summarised in Table~\ref{tab:method}.

\section{Innovation and survival}\label{sec:inno}

A key empirical stylised fact in industrial dynamics is the widespread heterogeneity of firms along many dimensions ranging from firms distribution of size, productivity, their growth, and chance of survival \citep{griliches1995production}. While this stylised fact clashes against the mainstream economics narrative of the representative agent, it is fully aligned with an evolutionary economics framework in which the mechanism of selection among heterogeneous agent explains the development of industries. In the last decades, scholars in this field made incredible progresses in understanding the driving forces which trigger selection and determine which firms prosper and succeed and which, on the contrary, fail. In particular, a paramount attention has been given to the entry of new firms and  how selection shapes their chance of survival and subsequent growth. More specifically, scholars focus on the role of innovativeness of new firms and whether innovation can explain, or at least improve, the fitness in the evolutionary landscape improving both survival and growth rate.

\subsection{Survival}
This literature is rooted in the Schumpeterian idea of creative destruction generated by the entry of new firms \citep{schumpeter1912theorie, schumpeter1942socialism} which are more prone to catch both market and technological opportunities than a large incumbent since they are not locked-in in partly obsolete competencies \citep{malerba1997, breschi2000technological}. For this reason, we mainly observe successful INNs in sectors with a high level of technological opportunity and where the cumulativeness of knowledge is low, that is previous competencies are not a strategic asset but more likely a burden which hinder the possibility of exploiting new ideas (ibid.). 
This Schumpeterian view shaped the mythological idea of the entrepreneur and serial entrepreneur, who challenges the odd of the fortune and after many attempts eventually succeed. When new innovative firms succeed, they grow fast like gazelles \citep{acs2008employment} and, in few case, they also became rare unicorns \citep{simon2016catch}.

However, this narrative is not always backed  by empirical facts. There are both clear advantages and disadvantages in pursuing an innovative ventures. Innovative firms might introduce better products and services which can improve users and consumers utility \citep{guerzoni2010impact}, they are less myopic and can focus on emerging markets \citep{bower1995}, they have less cognitive biases generated by previous activities \citep{aestebro2007}, and they are more dynamic \citep{teece2012}. At the same time, there is a high degree of uncertainty which can undermine their innovative efforts and bring them quickly to failure. In new markets there exists a high uncertainty about consumers' preferences \citep{guerzoni2010impact} and about the future development of the technology \citep{dosi1982}; there exists also an uncertainty due to competition, since other firms might win the race and take the lead of the market \citep[among others]{fudenberg1983}; it might be more difficult to find investments \citep{stucki2013success}.
Lately, scholars are forming a consensus which suggests a survival premium for innovative firms. However, this consensus does not seem to be rooted in a strong empirical evidence. Consider for instance the very precise review  on this issue by \citet{hyytinen2015does}, who survey the most relevant works \citep[][to mention a few]{arrighetti1999role, audretsch1995innovation, calvo2006testing, cefis2005matter, cefis2006survivor, colombelli2016, helmers2010innovation, santarelli2007entrepreneurship, wagner2010patents}. They classify empirical works according to the sign of the effect of innovativeness on survival probability, specify sample and proxy used for measuring firms' innovativeness and, eventually, conclude that the large majority of the works account for a positive impact. A rigorous reading of the paper shows however that the evidence of positive effects is rather weak.  For instance, \citet{cefis2005matter} find a close to zero effect, while \citet{cefis2006survivor}, although reporting a more robust result, do not control for the sector. In a very detailed work, \citep{colombelli2013} showed that the Kaplan-Meier survival function is virtually the same for innovators and non-innovators, while on a sample of French start-ups \citet{colombelli2016} show  that being innovative is not enough to have better survival chances than non innovative firms, and yet a very small effect on survival emerges for process innovation only. \citet{helmers2010innovation} use  patent activity as a proxy for innovation and find a mild positive and significant effect of patenting on survival, but due to the large sample the simple use of p-value can not really highlight anything conclusive. Indeed, when they repeat the analysis at the industry level and, hence, with less observations for regression, coefficients are still negligible in size and the p-value is significant for some sectors only. On a sample of U.S. listed companies, \citet{wagner2010patents}  find very small impact of owning patent on firm survival and coefficients are also mildly significant and for few specifications only. There also works included in the survey which show a negative impact of innovativeness on survival such as \citet{boyer2014born}. All in all, and just to mention a few, the works surveyed by \citet{hyytinen2015does} do not provide robust  evidence that ``[$\dots$] [T]the prevailing view in the empirical literature appears to be that there is a positive association between the innovativeness of firms and their subsequent survival'' (ibid. p.12). 
    
We highlight three main issues with the present empirical literature which might explain the disparate effect of innovation on the performance of entrants \citep{audretsch1995innovation}.

\paragraph{The measurement of innovation}
The research community of innovation studies has always acknowledged a number of shortcomings in the measurement of innovation,  but this is  rarely addressed in empirical works and mostly relegated in footnotes. Even the Oslo Manual 2018 \citep{manual} spends just a few words on the limits on the measurement of innovation that we would like to recall in the next paragraph.
The proxies adopted in empirical research for measuring innovation can be roughly divided in two groups: proxy for innovation input and proxies for innovation output. The input of the process of innovation are typically R\&D investments and high-skilled labour, while, as for as the innovation output, the number of product or process innovation or patent application.
 Figure on R\&D expenses and personnel costs come usually from register data, patents are easily identified in patents office and there is an extensive literature on their uses, while information about the number and nature of new products or processes can be found in self-reported surveys such as the Community Innovation Survey.

Each of these empirical proxies proved to have important downsizes, which are even more severe for recently established firms. R\&D expenses in register data are not always representative of real R\&D activity especially in small enterprises for which R\&D is not pursued in a formal way or in high-tech start-ups for which, conversely, the R\&D activity is spread out across any firm operation. The number of product and process innovations are biased towards the misrepresentation of the concept of innovation of the respondent \citep{manual}.
Moreover, surveys do not cover all the population of firms, typically start-ups,  and, thus, the process of sample selection can induce bias, reduce the possibility of panel data, reduce the degree of freedom of the model and, thus, raise problems with the inference.
As for patents, there is clear evidence on the extreme variance in the propensity to patent both between and within sectors, since in many cases, especially for process innovation, appropriability of the economic returns of IPR can be achieved by mean of secrecy \citep{harabi1995appropriability}. In addition to that patents are an indicator of the inventing activity and only rarely they turn to be commercially valuable since the patenting activity is pursued for a vast array of purposes\footnote{The debates on the use of patent dates back at least to the work by \citet{pavitt1985patent}.}.

These measurement issue are even more stringent for start-ups since the balance sheets in the first years are rarely a precise representation of the firm business and, as for patents, start-ups might still be in the application process or decide not-to-patent since in some contexts  time-to-market might be much more important than a strong IPR.

\paragraph{Business cycles as confounding effect}
New firms can prosper or fail for a large variety of factors which do not necessarily relate with economic or technological conditions at the micro level. For instance, vulnerable firms might survive in a growing economy even if not profitable, while selection mechanisms become stricter in downturns. The literature on economic and financial crisis agrees that recessions usually hinders survival  for existing firms. \citet{peric2016} review the literature of the adverse effect of crisis on existing firms and highlight the main channels such as  production and product lines \citep{liu2009}, sales \citep{cowling2014}, employment \citep{rafferty2013}, investments \citep{campello2010, buca2012}, performance \citep{akbar2013impact}, risk tolerance \citep{hoffmann2012, inklaar2012} and business confidence \citep[][p.3]{zenghelis2012, geels2013, peric2016}. However, entrepreneurial studies has also stressed positive effects produced by an economic crisis \citep{bartlett2008}. This is especially true for those firms that can identify  changes in the market and react promptly to exploit new opportunities \citep{hodorogel2009}. For this reason, if there are clear differences in firms survival growth between innovative and non-innovative firms, we should be able to spot them more neatly from this cohort of firms born in 2008, when business constraints became more binding. 

\paragraph{Sectors and location as confounding effects}
Since the work by \citet{pavitt1984sectoral}, it has been widely acknowledged that the sector specificity play a crucial role in explaining the performance of a sector especially in terms of innovation. Along the same line, the work by \citep{malerba1997} developed a theory and provided strong empirical evidence that the technological base underlying the activities of a sector is a key driver of the innovative performance of firms. Sectors characterised by high technological opportunities, low appropriability and a low cumulativeness of the technological knowledge experience a high entry rate of innovative firms, but also a high rate of exits.
Along the same line, the industry life cycle approach theorised and showed that the early stages of new industries attract most of the entries, but at the same experience the highest rates of failures \citep{klepper1996entry, geroski1995we}. Within an evolutionary perspective this can be framed as the costly process of trials and errors  at the industry level in which many enter, but the most do not survive: survivors thereafter exhibit a more than proportionate growth base on their performance (ibid.). 
 Thus, \citet{pavitt1984sectoral}'s taxonomy, \citet{malerba1997}'s classification,  and \citet{klepper1996entry}'s industry life cycle approach suggest that there exist a bias for innovative firms towards specific sectors and survival rates might differ between innovative and non innovative firms because of a self-selection of innovative firms in specific sectors with specific patterns of survival.

Similarly, since the distribution of economic activities is very uneven across space, regions' specific fixed effects can introduce a further confounding effect when analysing  the survival rate.  The impact of a region on the economic performance is heavily determined by the spatial distribution of economic activity at the industry level, however \citet{acs2007determinants} show that even after controlling for both the industry mix of an area and its degree of specialisation there is still an effect of location on survival. 
As  \citet{sternberg2004regional, sternberg2009regional} recall and show that entrepreneurship is mainly ``a regional event'' \citep{feldman2001entrepreneurial} for many other reasons which can be broadly define as agglomeration economies \citep{leone1976, sorenson2000}, the regional system of innovation \citep{howells1999}, which might include among others local government policies, specific user-producer interactions\citep{rothwell1994}, the presence of an entrepreneurial atmosphere \citep{ciccone1996}, the role of cities \citep{lee2004}, industrial clusters \citep{rocha2004}, and the presence of high tertiary education institution or research centres \citep{fetters2010}:  knowledge spillovers are the key input in the complex process of innovation especially for new entrants \citep{audretsch1996r, audretsch2004knowledge}.
First, there is not a consistent use of the control for industries and regions which the theoretical literature suggested as the most important. For instance, none of the work discussed \citep[among others][]{arrighetti1999role,audretsch1995innovation, calvo2006testing, cefis2005matter, cefis2006survivor, colombelli2016, helmers2010innovation, santarelli2007entrepreneurship, wagner2010patents} controls for the location and not all of them control for the sector.

Both theoretical and empirical considerations trigger the necessity of a novel approach to the problem of survival. In this paper we aim at providing a solution to three issues discussed above and therefore, we look for evidence of different survival rates between INNs and NOINNs by (i) introducing a new empirical measure of innovativeness, clearly, (ii) by focusing on the population of new Italian firms in time of crisis, and (iii) controlling for sectors and location as suggest by the theory.

The contextual achievements of these three goals  pulls the necessity of developing a challenging methodology which is the main contribution of this paper. More in details, we
\begin{itemize}
    \vspace{-0.2cm}\item provide new evidence on the survival;
    \vspace{-0.2cm}\item provide a new way to detect innovative firms with a scope larger than the simple question about survival;
    \vspace{-0.2cm}\item provide a methodological framework to combine data science and econometrics.
\end{itemize}

\section{Data and methodology} \label{data}

The AIDA (Analisi Informatizzata delle Aziende) database, provided by Bureau van Dijk, contains comprehensive information on all Italian capital companies required to fill their accounts, including whether they have registered as INNs or not according to the Decree-Law 179/2012 (then converted in the 221/2012 Law in force since 17 December 2012). 

Each  firm in AIDA is described by 427 variables belonging to the following macro categories: i) identification codes and vital statistics; ii) activities and commodities sector; iii) legal and commercial information; iv) index, share, accounting and financial data; v) shareholders, managers, company participation. Only variables in category iv) are observed for different years. In the construction of our star-ups sample, we excluded category v) since the nature of this data is very specific to each observation and not suitable for prediction analysis, nor for econometrics. Despite its considerable dimension (Table~\ref{tab:inn1}), the AIDA database does not cover the entire population of Italian firms and, for instance, banks, insurances and public bodies are not included. Still our sample varies from a minimum of 62,934 observations in 2009 (about 21.8\% of new firms) to a maximum of 74,508 in 2010 (about 28\% of new firms). 
The dataset collects 276 variables for all firms entering the Italian market from 2008 to 2015, out of which 262 are observed from the starting year of activity until 2015. For new Italian firms established in 2008, we have a balanced panel with ten selected variables up to 2018. Since not all information is mandatory for each category of firms, the dataset is characterised by many missing values. Therefore, we conduct a careful missing value analysis which brought us to exclude some variables and observations and obtain two samples respectively of 45,576 (2013) and 39,295 (2008) observations.  Appendix~\ref{mva} includes details on our cleansing methodology.

\subsection{Measuring innovation: innovative start-ups and Law 221/2012}\label{law}
In the previous section, we suggested that we cannot rule out the possibility that the weak evidence in the empirical literature on survival and innovation depends on measurement issues. Usually, the literature assesses the innovativeness of firms looking either at inputs of the innovation process, such as R\&D expenses or number of employed researchers, or outputs of the process such as the patent pool or the number of innovations.
For this reason, we adopt a new definition of ``innovative start-ups'' which allows to grasp simultaneously several dimensions of their innovative activity.

Starting from the 2012, a new Italian class of firms named ``innovative start-up'' has been identified through the Decree-Law 179/2012, then converted into the Law 221/2012 in force since 17 December 2012.
The policy recognises the disadvantaged position of start-ups, but intends to encourage the creation of companies with specific characteristics, such as development, production or marketing of innovative products or services with high technological value. The Law applies also to firms already active in the market for a period less than four years from the adoption of the Law, however it is only since 2013 that this opportunity has been consistently exploited by firms. To identify the beneficiaries of the policy, the Law sets up a specific section in the Italian companies register\footnote{See website: http://startup.registroimprese.it/startup/index.html\label{foot:1}}. The registration allows for specific incentives at different levels for the first five years of activities: registration and fiscal incentives are tailored as well as a specific labour legislation for INNs in order to introduce a higher level of labour flexibility and a fail-fast procedure for firms.
Start-ups applying for these incentives must meet the following requirements:
\begin{itemize}
  \item be new or have been operational for less than five years;
  \item have their headquarters in Italy or in another EU country, but with at least a production site branch in Italy;
  \item have a yearly turnover lower than 5 million Euros;
  \item do not distribute profits;
 \item produce, develop and commercialise innovative goods or services of high technological value;
\item are not the result of a merger, split-up or selling-off of a company or branch;
\item be of innovative character, which can be identified by at least one of the following criteria:
\begin{itemize}
\item at least 15\% of the company's expenses can be attributed to R\&D activities (satisfied by 64.97\% of the INNs);
\item at least 1/3 of the total workforce are PhD students, the holders of a PhD or researchers; alternatively, 2/3 of the total workforce must hold a Masters degree (satisfied by 29.68\%of the INNs);
\item the enterprise is the holder, depositary or licensee of a registered patent (industrial property) or the owner of a program for original registered computers.
\end{itemize}
\end{itemize}

\noindent Accordingly to the actual composition of INNs, only the 2.7\% satisfied all the three requirements and the 11.08\% is characterised by two up to three requirements. From AIDA, we do not know which specific criteria they satisfy to be registered as innovative. We only have aggregate data from the Italian Board of Trade (IBT), presented in Table~\ref{tab:crit} for 2013. 

The Law 221/2012 provide us with a new tool to identify INNs with some advantages over previous indicators of innovativeness:

\begin{itemize}
\item we focus on small firms, which are very likely to be truly new entities and not subsidiaries or foreign green-field entrants;
\item all innovative firms are focused on innovative goods or services;
\item they need to have at least one of the usual proxy for innovative input and output, but not necessarily a specific one such as in the other measures.
\end{itemize}

\noindent Table~\ref{tab:inn1} shows the numbers of INNs over the total number of firms in the sample and the percentage of firms in data over the entire population (source: see Footnote~\ref{foot:1}) of new Italian firms. AIDA covers about a fifth of new Italian firms, since firms self-employer, professional and other minor activities are not required to fill their accounts. In 2013, firms registered as innovative start-ups are about the 1.5\%.

\begin{table}[htbp]
\caption{INNs and NOINNs in the collected sample accordingly to the initial year of activity. Source: AIDA}
\centering
\begin{footnotesize}\begin{tabular}{lrrrrr|r}
  \hline  \hline
 & 2008 & 2009 & 2010 & 2011 & 2012 & 2013  \\ 
  \hline  
INNs & 0 & 4 & 51 & 320& 531 & 1,010  \\ 
NOINNs & 65,088 & 62,930 & 74,457 & 71,599 &65,653 &67,306\\ 
   \hline  
   Total&65,088	&62,934&	74,508	&71,919&	66,184	&68,316	 \\ 
\% Italian Start-ups (IBT)&22.7\%	&21.8\%	&28.1\%	&27.2\%	&24.0\%	&24.7\%\\ 
   \hline  
   \% INNs (AIDA) &0\%  & 0.01\% & 0.07\% &  0.4\% & 0.8\% & 1.5\%\\ 
   \% INNs w.r.t. population (IBT) & 0\% & 21.05\% & 43.22\% &  124.51\% & 112.03\% & 100.50\%\\ 
\hline\hline
\end{tabular}\end{footnotesize}
\label{tab:inn1}
\end{table}

\begin{table}[htbp]
\caption{Number of 2013 INNs satisfying the three possible criteria. Source: IBT, march 2016}\centering
\begin{footnotesize}\begin{tabular}{rrrr}
  \hline  \hline
 & First Req.  & Second Req.  & Third Req.\\ 
  \hline  
No & 360 (35.82\%) &709 (70.55\%)&787 (78.31\%)\\ 
Yes & 645 (64.18\%) &296 (29.45\%) & 218 (21.70\%)\\ 
   \hline  \hline
\end{tabular}
\end{footnotesize}
\label{tab:crit}
\end{table}


 \begin{table}[htbp]
\caption{One digit ATECO2007 of the 2013 INNs ($n=1,010$) and NOINNs (67306). Source: AIDA}\centering
\begin{footnotesize}
\begin{tabular}{ccc}
  \hline  \hline
 ATECO & NOINNs& INNs\\
& start-ups& start-ups\\ 
  \hline 
  A & 1,039 (1.54\%) &6 (0.59\%) \\ 
  B &  46 (0.06\%)&0 (0.00\%)\\ 
  C & 7,112 (10.56\%) &161(15.94\%)\\ 
  D & 703 (0.10\%)&10 (0.10\%)\\ 
  E & 326 (0.48\%)&4 (0.39\%)\\ 
  F & 8,290 (12.32\%)&14 (1.39\%)\\ 
  G & 15,415 (22.90\%) &59 (5.84\%)\\ 
  H & 2,640 (3.92\%) &3 (0.30\%)\\ 
  I & 6,072 (9.02\%)&0 (0.00\%)\\ 
  J & 3,113 (4.63\%)&431 (42.67\%)\\ 
  K & 1,309 (1.94\%)&1 (0.10\%)\\ 
  L & 3,193 (4.74\%)&0 (0.00\%)\\ 
  M & 4,963 (1.54\%)&261\\ 
  N & 4,260 (7.37\%)&37(3.66\%)\\ 
  O &   6 (0.00\%)&0 (0.00\%)\\ 
  P & 666 (0.99\%)&7 (0.69\%)\\ 
  Q & 1,307 (1.94\%)&6 (0.59\%)\\ 
  R & 1,865 (2.77\%)&3 (0.30\%)\\ 
  S & 1,098 (1.63\%)&4 (0.40\%)\\ 
  T &   1 (0.00\%)&0 (0.00\%)\\
  \hline
  Total &67,306 & 1,010\\
   \hline  \hline
\end{tabular}
\end{footnotesize}
\label{tab:ateco}
\end{table}

\noindent Table~\ref{tab:ateco} describes the distribution of INNs across the ATECO2007 sector classification and  shows that INNs are principally active in service and manufacturing (code J and C, respectively). The geographic map of 2013 INNs (see Table~\ref{tab:innov3} and Figure~\ref{fig:1} in Appendix~\ref{app:descr}) shows a striking concentration in two regions, Lombardia and Lazio, and their capital cities, Milan and Rome, which attract about one out of three INNs. 

\begin{table}[htbp]
\caption{Regional distribution (NUTS2) of the 2013 INNs ($n=1,010$). Source: AIDA }\centering
\begin{footnotesize}
\begin{tabular}{lrlr}
\hline\hline
 Italian Region& Number of innovative &Italian Region& Number of innovative  \\
 & start-up&& start-up \\
  \hline
  Abruzzo &  16 (1.58\%)&Molise &   0 (0.00\%)\\ 
  Basilicata &   4 (0.40\%)&Piemonte (Torino - 57)&  72 (7.13\%)\\ 
  Calabria &  14 (1.39\%) &  Puglia &  51 (5.05\%)\\ 
  Campania &  65 (6.44\%)&  Sardegna &  30 (2.97\%)\\ 
  Emilia-Romagna & 111 (10.99\%)&  Sicilia &  39 (3.86\%)\\ 
  Friuli-Venezia Giulia &  26 (2.57\%)&  Toscana &  53 (5.25\%)\\ 
  Lazio (Roma - 102)& 111 (10.99\%)&  Trentino-Alto Adige &  27 (2.67\%)\\ 
  Liguria &  15 (1.49\%) &  Umbria &  13 (1.29\%)\\ 
  Lombardia (Milano - 155)& 229  (22.67\%)&  Valle d'Aosta &   3 (0.30\%)\\ 
  Marche &  47 (4.65\%)&  Veneto &  84 (8.32\%)\\ 
   \hline\hline
\end{tabular}
\end{footnotesize}
\label{tab:innov3}
\end{table}

\noindent Finally, we conclude the presentation of INNs main features proposing some summary tables on the activity state (the 98\% is still active in the 2015, see Table~\ref{tab:innov5}). It is here to notice that in the survival analysis we will consider as a firm's death only the negative exits, such as closing or failures.

\begin{table}[htbp]
\caption{Activity state of the 2013 INNs ($n=1010$). Source: AIDA }\centering
\begin{footnotesize}
\begin{tabular}{lrrrrrrr}
\hline\hline 
Status &Number of INNs\\
\hline
Active & 994 (98.42\%)\\ 
  Close down &   0 (0.00\%)\\ 
  Failed &   1 (0.1\%)\\ 
Liquidation &  15 (1.49\%) \\
     \hline\hline
\end{tabular}
\end{footnotesize}\label{tab:innov5}
\end{table}

\subsection{Isolating the innovators' premium from confounding effects}

Unfortunately for the purpose of this paper, which aims at classifying and studying the survival of firms born in 2008, the Law was introduced in 2012 and only since 2013 consistently exploited by new firms. 
In this paragraph, we explain how a machine learning algorithm can be trained and tested on 2013 data to identify INNs in 2008 without specific information on model assumptions, but based on a vast array of other firmographics. We can use any type of variable, with the only restriction provided by the requirements of the theory and the econometrics model to be performed. Namely, we excluded from the analysis variables related with the industry sector and the geographical location, otherwise, the new indicator would have not been suitable as a regressor in a model in which sector and location appear as other covariates. At the same time, we also discard from the training-set the investment in R\&D and IPR. This exclusion will serve us later as  an evaluation tool for prediction in 2008 for which we do not observe the target variable, as explained in Section~\ref{sec:pred}. We focus on 2013 data since it is the closest available year with the relevant information about the target variable (INNs). Predictive modelling learning from historical data is assumed to be static, but data evolves and must be analyzed in near real time. The change over time of the statistical properties of the target variable, which the model is trying to predict, is also known as concept drift \citep{zlio2010}. Therefore, to prevent deterioration of the prediction accuracy, one effective solution is to minimise the time interval between input and output data.

\subsubsection{Training, test, and model selection to predict INNs}\label{sec:tt}

In this section, we apply different algorithms to classify firms as INNs and, thereafter, we compare their predictive power to select the most performing one. We have deployed seven widely used classifiers, which are analytically describe in Appendix~\ref{algo}:
\begin{itemize}
  
 \vspace{-0.2cm}\item Recursive Partitioning (RPART);
 \vspace{-0.2cm}\item Classification Tree (TREE);
 \vspace{-0.2cm}\item Conditional Inference Tree (CTREE);
 \vspace{-0.2cm}\item Bagging (BAG);
 \vspace{-0.2cm}\item Logit Regression (LOGIT);
 \vspace{-0.2cm}\item Na\"ive Bayes (NB);
 \vspace{-0.2cm}\item Artificial Neural Network (ANN).
\end{itemize}

\noindent We train these algorithms on a 2013 random subset of 80\% of the cleansed sample (36,401 firms including 563 INNs) and we test them on the remaining 20\% of the sample (9,175 firms including 150 INNs). The dataset is unbalanced since the target variable (INNs) is underrepresented in the samples. The SMOTE algorithm \citep{chawla2002} is a well-known technique to address this problem because it artificially generates new examples of the minority class (here INNs) using the nearest neighbours of these cases. Furthermore, the majority class examples are also under-sampled, leading to a more balanced dataset. Eventually, we synthetically increase the number of INNs cases in the training-set only, while we keep the test-set unchanged to evaluate the performance on factual data.

\begin{figure}[htbp!]
\begin{center}
\centerline{\includegraphics[scale=0.5]{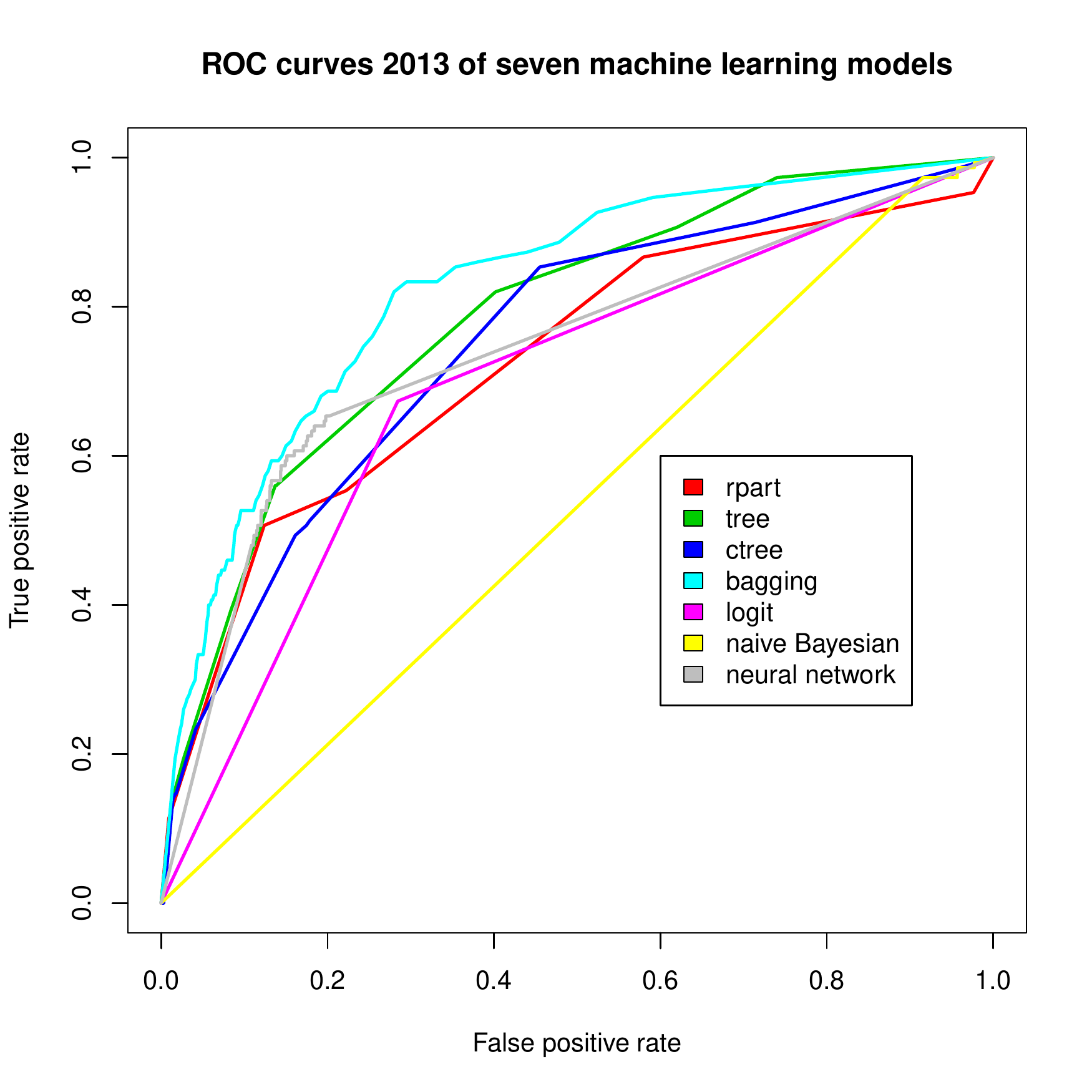}} \caption{ROC curves on the 2013 start-ups and given the SMOTE technique}\label{fig:roc2}
\end{center}
\end{figure}

Each algorithm predicts the probability of a start-up in test-set to be INN. The predicted probability, which maps from 0 to 1, collapses to NOINN or INN according to a threshold (or cut-off) chosen by the researcher on the basis of a model performance assessment. Our toolbox for comparing algorithms and selecting thresholds includes the analysis of receiver operating characteristics (ROC) curves (Figure~\ref{fig:roc2}) and the density function of the predicted probability for both INNs and NOINNs.
In details, the ROC curves represent pairs of true positive and false positive rates of a classifier for a continuum of probability thresholds and they can be used to compare different classifiers. Specifically, the highest performing classifier is the one with the ROC curve closest to the upper-left corner (\emph{i.e.} true positive rate close to 1 and false positive rate close to 0). If two classifiers are characterised by intersected ROC, it means that the two classifiers are better under different loss conditions\footnote{Alternatively, as measure of performance, we can compare the area under the ROC curve (AUC). For further details on the interpretation of ROC curves see \citet{Alpaydin}.}. For each algorithm, we define the optimal threshold (or cut-off) as the one associated with the point minimising the Euclidean distance between the ROC curve and the (0,1) point (see Appendix~\ref{cutofs} for further details). Once a cut-off is set, confusion matrices (Table~\ref{tab:perf2}) summarise the number of correctly classified cases and classification errors for each algorithm.

\begin{table}[htbp]
\caption{Confusion matrix of the seven algorithms with SMOTE. Optimal cut-offs in parentheses}\centering
\begin{footnotesize}
\begin{tabular}{l|cc|cc|cc|cc|l}
\hline\hline
\multirow{3}{*}{Real Data}                      & \multicolumn{2}{c|}{RPART}    & \multicolumn{2}{c|}{TREE}     & \multicolumn{2}{c|}{CTREE}    & \multicolumn{2}{c|}{BAG}      & \multirow{3}{*}{Total} \\
                                                & \multicolumn{2}{c|}{(0.2817)} & \multicolumn{2}{c|}{(0.3210)} & \multicolumn{2}{c|}{(0.2632)} & \multicolumn{2}{c|}{(0.1200)} &                        \\
                                                & NOINNs          & INNs        & NOINNs         & INNs         & NOINNs          & INNs        & NOINNs         & INNs        &                        \\ \hline
NOINNs                                          & 7,908            & 1,117        & 7,779           & 1,246         & 6,857            & 2,168        & 6,612           & 2,413        & 9,025                   \\
INNs                                            & 74              & 76          & 66             & 84           & 42              & 108         & 32             & 118         & 150                    \\
                                                &                 & 50\%        &                & 56\%         &                 & 72\%        &                & 79\%        &                        \\ \hline
\multicolumn{1}{c|}{\multirow{3}{*}{Real Data}} & \multicolumn{2}{c|}{LOGIT}    & \multicolumn{2}{c|}{NB}       & \multicolumn{2}{c|}{ANN}      &                &             & \multirow{3}{*}{Total} \\
\multicolumn{1}{c|}{}                           & \multicolumn{2}{c|}{(1)}      & \multicolumn{2}{c|}{(1)}      & \multicolumn{2}{c|}{(0.1905)} &                &             &                        \\
\multicolumn{1}{c|}{}                           & NOINNs          & INNs        & NOINNs         & INNs         & NOINNs          & INNs        &                &             &                        \\ \hline
NOINNs                                          & 9,025            & 0           & 9,023           & 2            & 7,240            & 1,785        &                &             & 9,025                   \\
INNs                                            & 150             & 0           & 150            & 0            & 53              & 97          &                &             & 150                    \\
                                                &                 & 0\%         &                & 0\%          &                 & 65\%        &                &             &                       
\\\hline\hline\end{tabular}\label{tab:perf2}
\end{footnotesize}
\end{table}

\begin{figure}[!tbp]   
\def\tabularxcolumn#1{m{#1}}
%
\begin{tabular}{cc}

    \subfloat[WELL BEHAVED]{\includegraphics[scale=0.37]{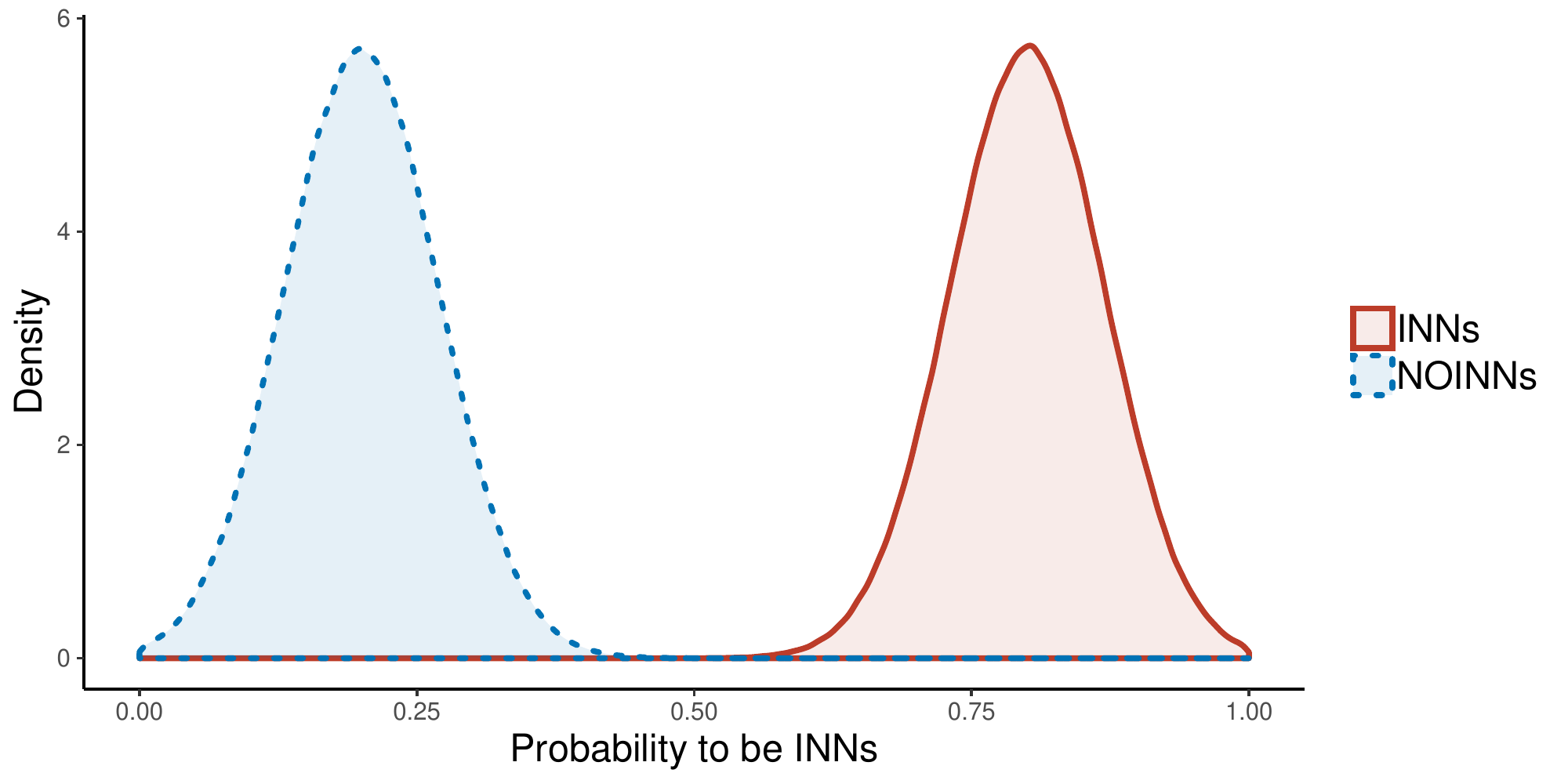}\label{fig:wellB}}

  &
  \subfloat[RPART]{\includegraphics[scale=0.37]{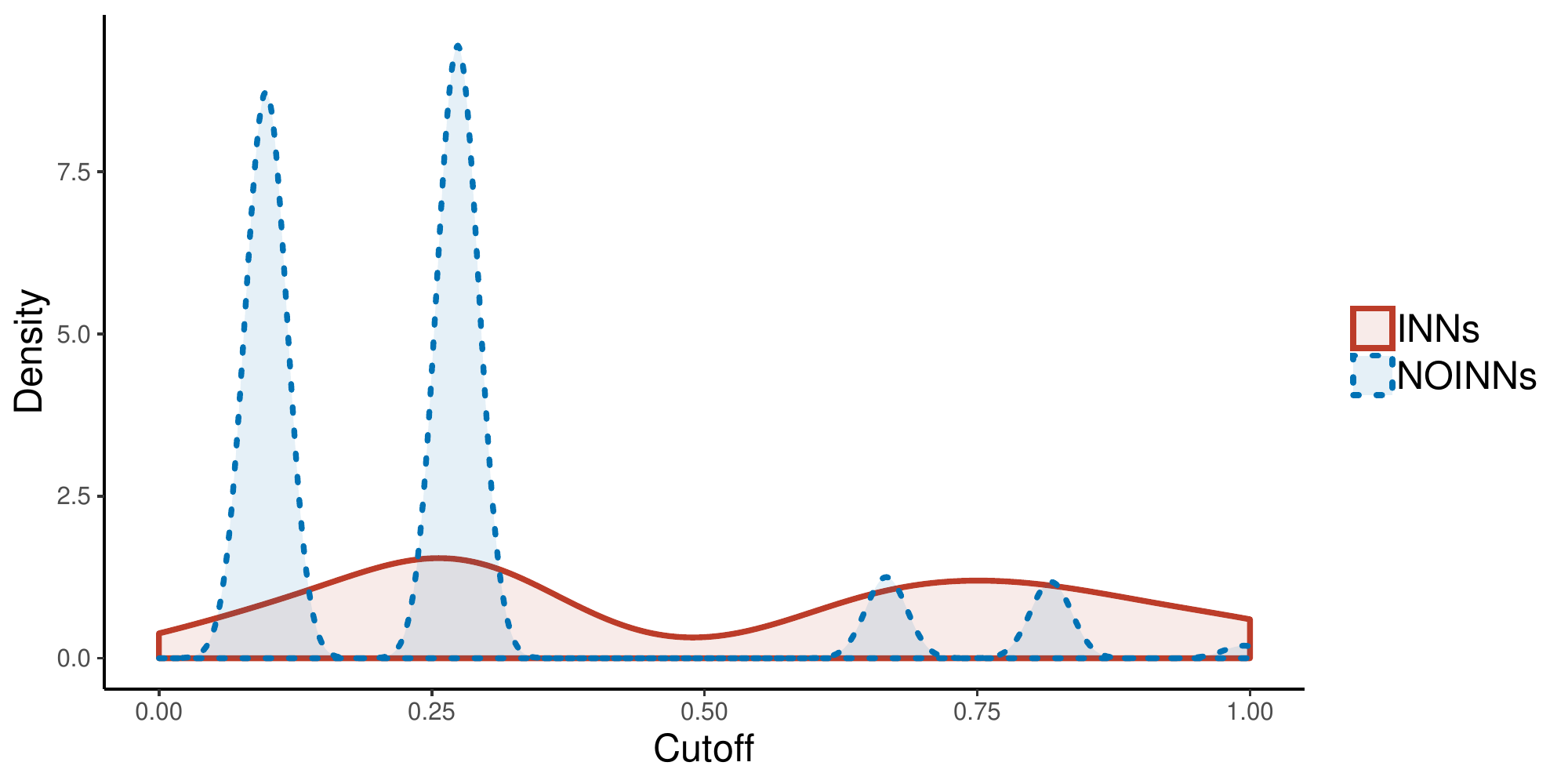}\label{den:f1}}\\

  \subfloat[TREE]{\includegraphics[scale=0.37]{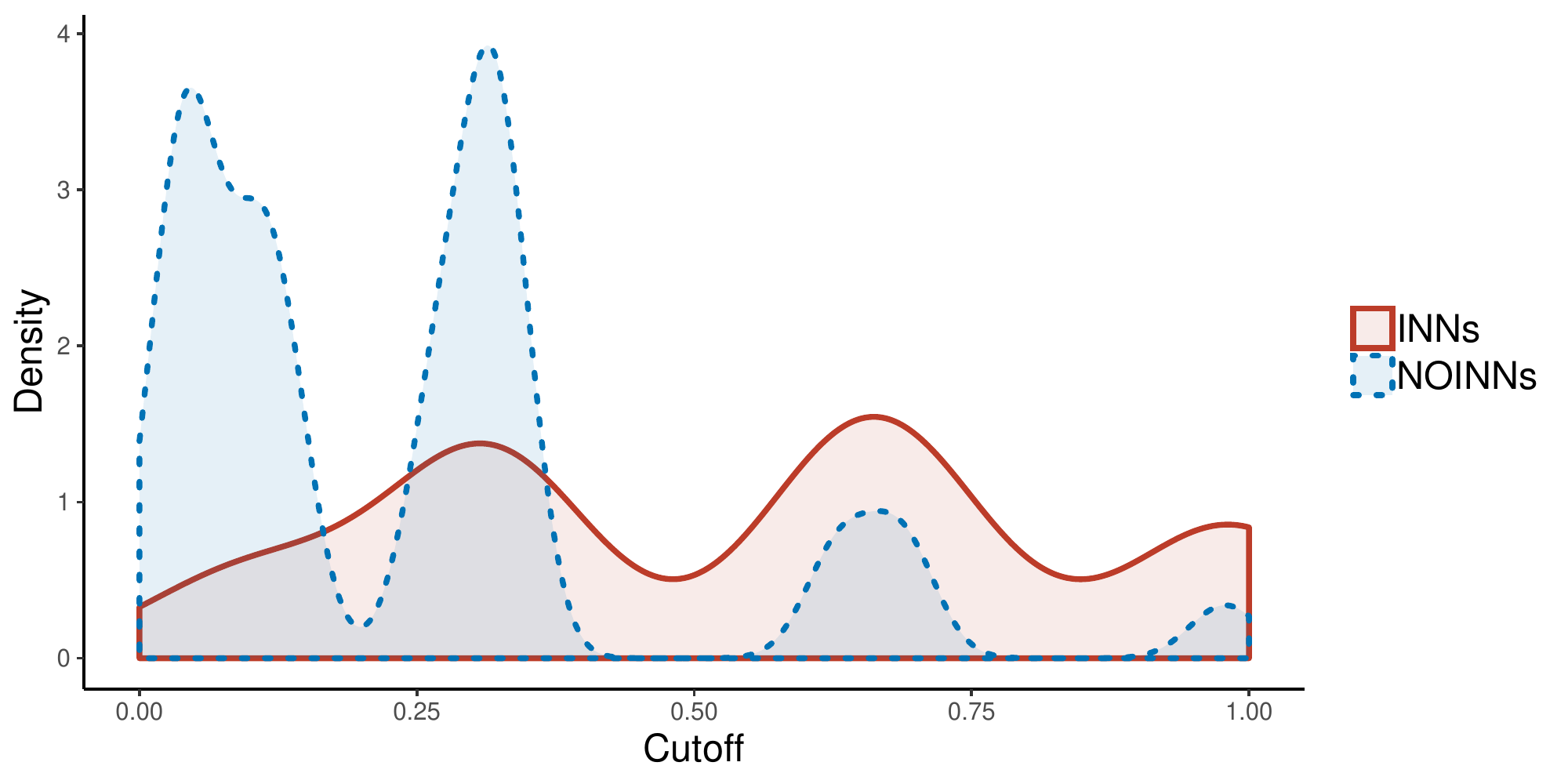}\label{den:f2}}
  &
    \subfloat[CTREE]{\includegraphics[scale=0.37]{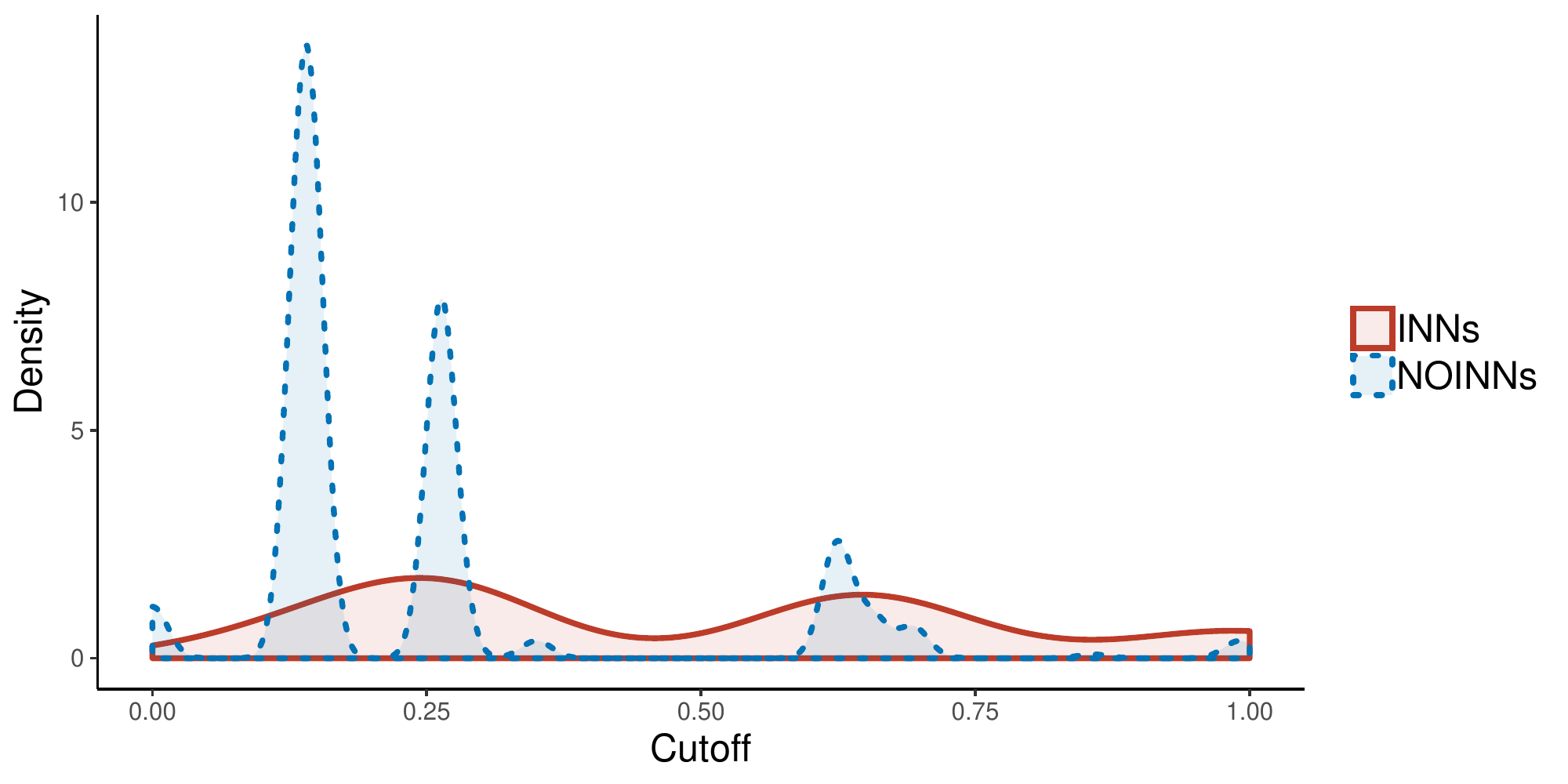}\label{den:f3}}\\

  \subfloat[BAG]{\includegraphics[scale=0.37]{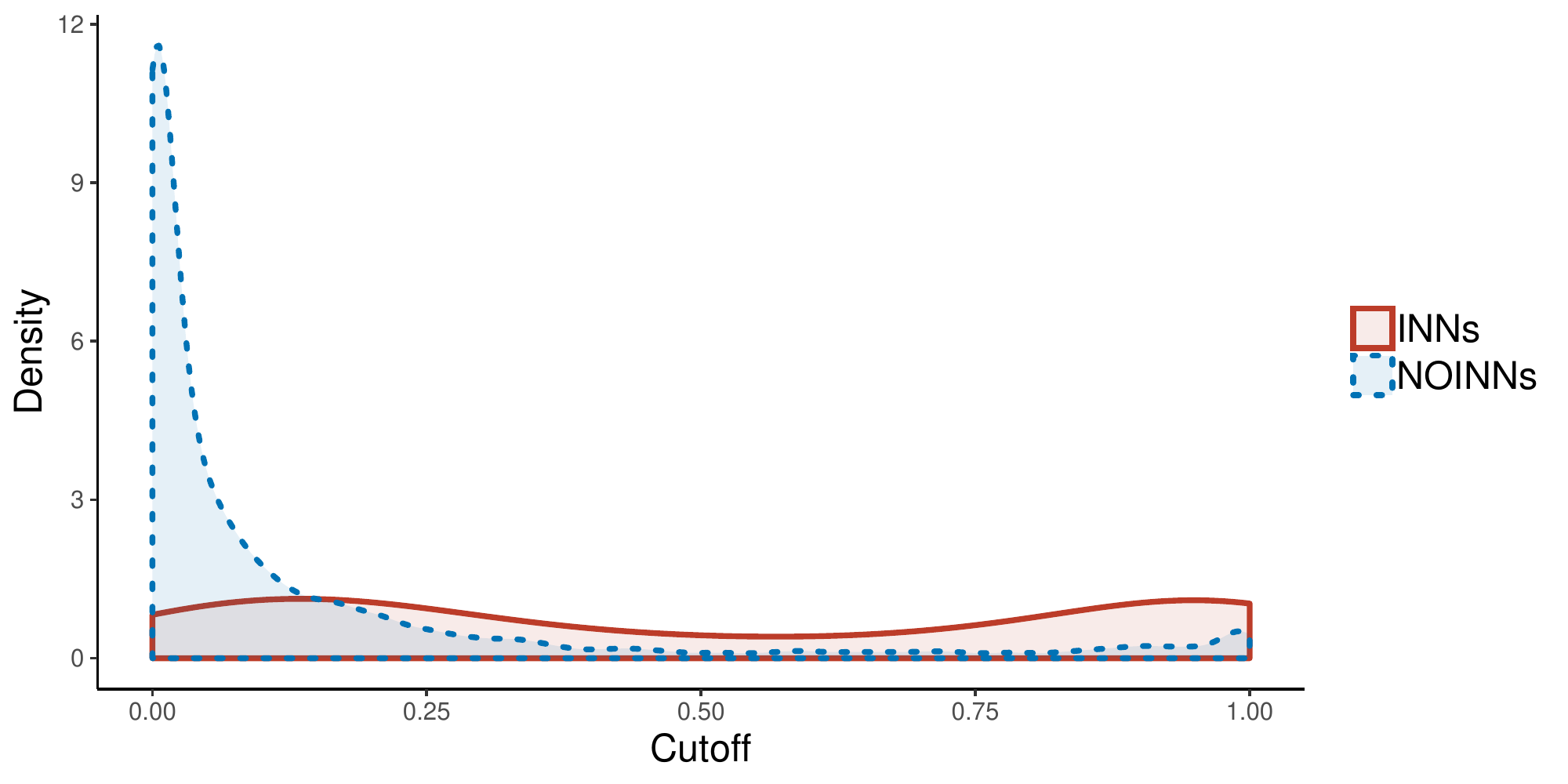}\label{den:f4}}
  &
    \subfloat[LOGIT]{\includegraphics[scale=0.37]{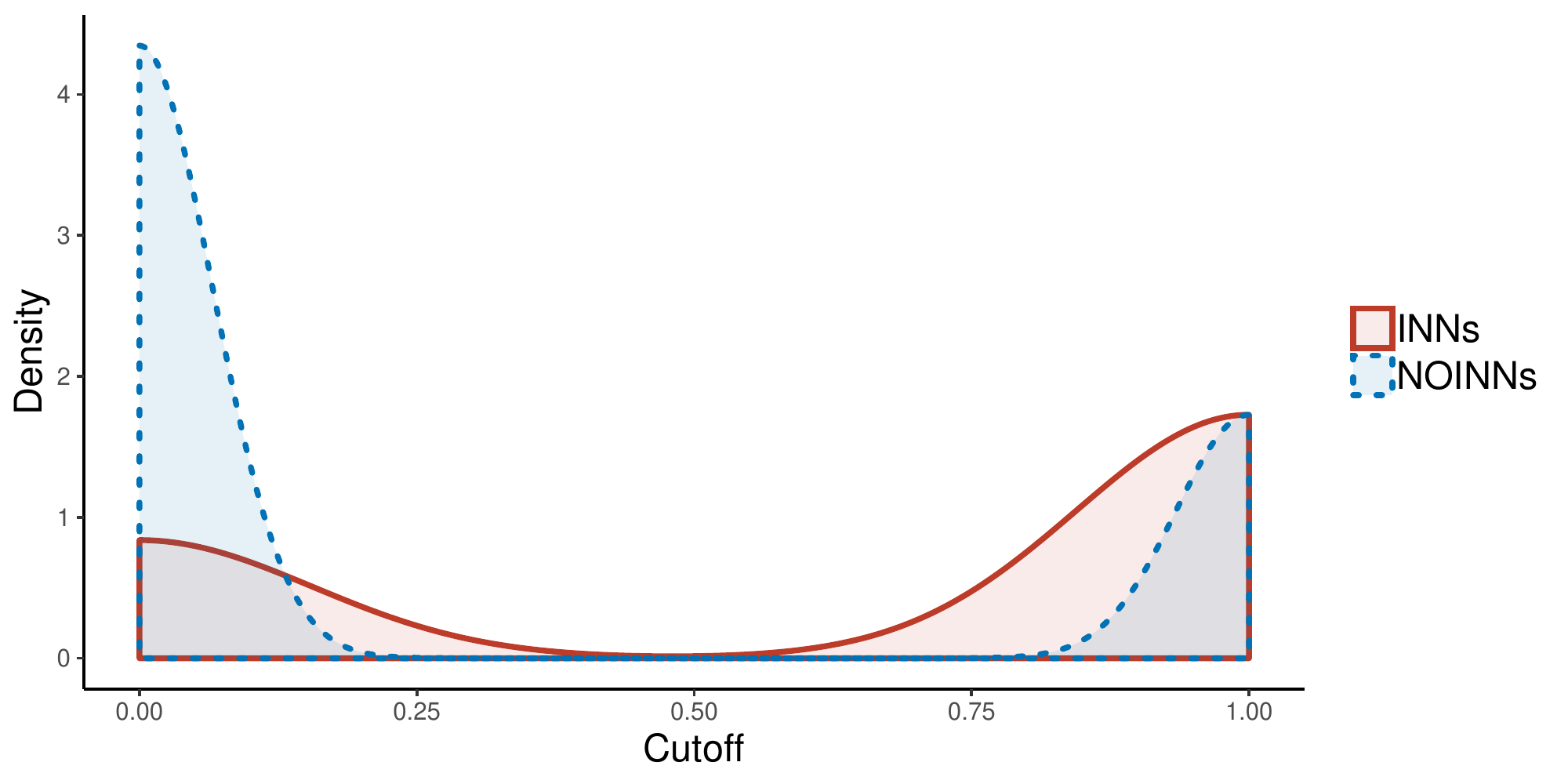}\label{fig:f5}}\\
  \hfill
  \subfloat[NB]{\includegraphics[scale=0.37]{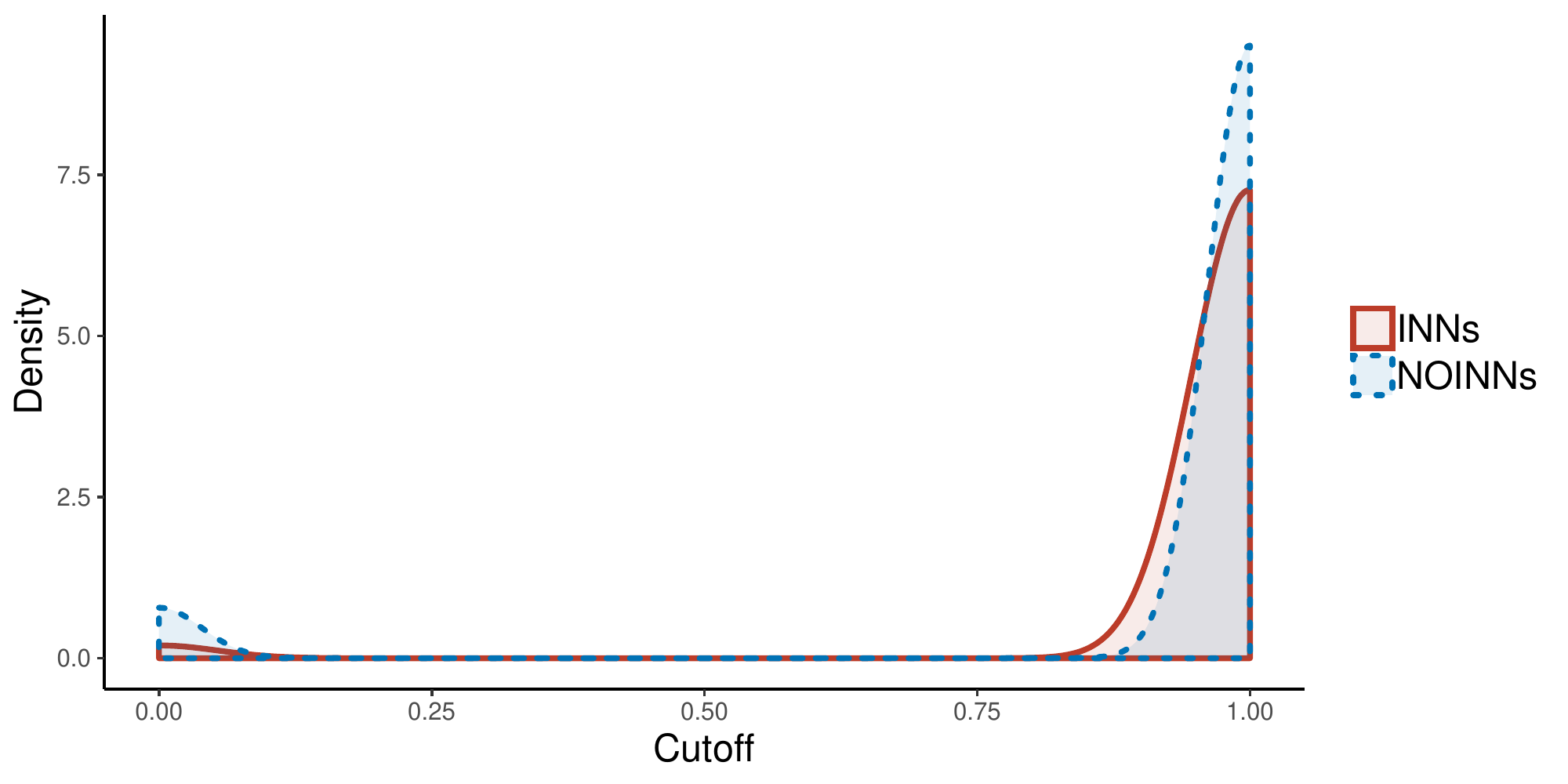}\label{den:f6}} 
  &
  \subfloat[ANN]{\includegraphics[scale=0.37]{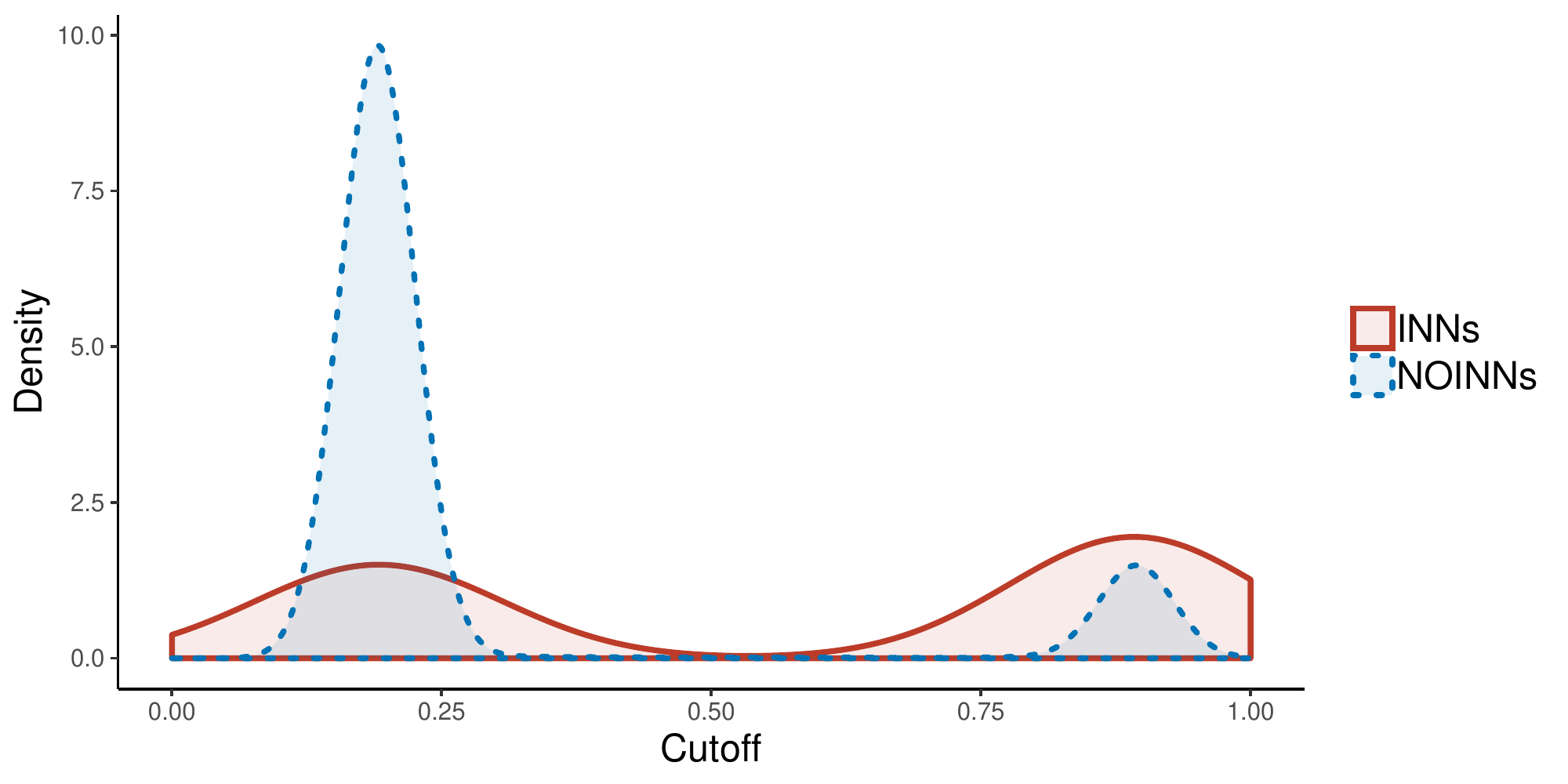}\label{den:f7}}
  \hfill
  \end{tabular}
\caption{INNs and NOINNs predicted probabilities for the 2013 start-up according to the seven machine learning algorithms with SMOTE and the well-behaved example, as a benchmark}  \label{density}

\end{figure}

Also the density function for both NOINNs (or negative) and INNs (or positive) predicted probabilities (Figure~\ref{density}) generated by the seven algorithms on the 2013 test-set can provide some insights on the model performance. In the ideal scenario, represented in Figure~\ref{fig:wellB}, each distribution for the predicted probability of INNs and NOINNs shall be skewed respectively towards 1 and 0 and without a common support. Unfortunately, this is not the case and, in most empirical analyses, I and II type misclassification errors can be relevant.

\begin{figure}[htbp]
\begin{center}
\centerline{
 \includegraphics[scale=0.5]{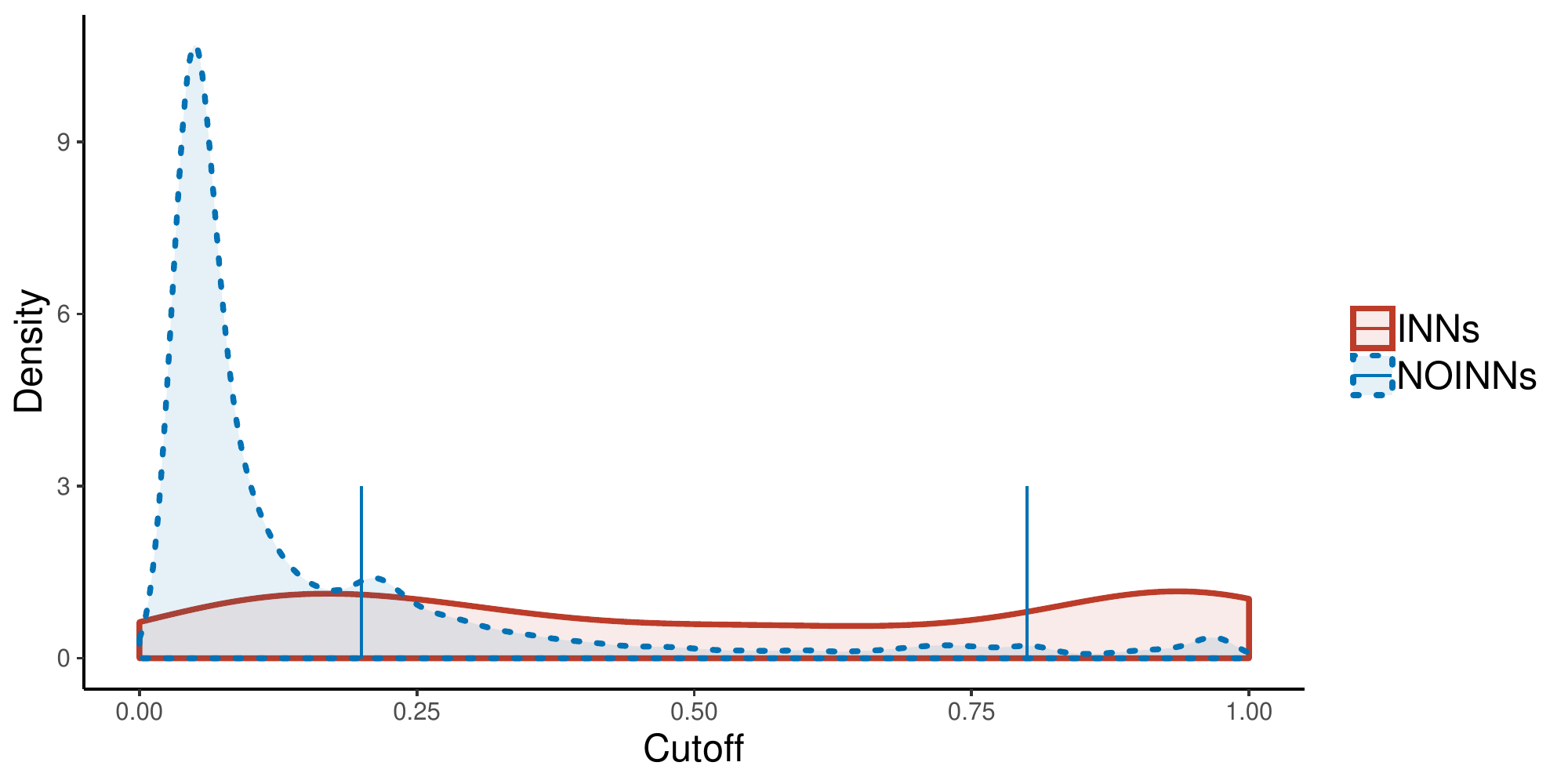}} \caption{Density of the predicted probabilities for the 2013 start-ups accordingly to the BAG-ANN mixture with SMOTE}\label{fig:dens5}
\end{center}
\end{figure}

 ROC curves offer an interesting comparative insight on the performance of the same model when used for estimating a causal relation instead of predicting a category, as discussed in Section~\ref{sec:bigdata}. Indeed, Figure~\ref{den:f4} shows how the Logit model, well-known in econometrics and here used to classify start-up, is one of the worst classifiers. This confirm the theoretical framework portrayed in Section~\ref{sec:bigdata} and main limits of econometrics used to fit such type of data.
Among the ones tested in this exercise, the best predicting algorithm (with SMOTE) is the BAG. When considering the optimal cut-off in the 2013 sample (0.12), this algorithm classifies 6,644 to be NOINNs and 2,531 (38.1\%) to be the INNs. 
Unfortunately, the predicted probability distributions associated with BAG does not identify correctly many INNs (see Figure~\ref{den:f4}). The distribution of the predicted probabilities for NOINNs is well-behaved while for INNs is bi-modal with a large variance across its domain. For this reason, the common support is the domain itself. The second best performing algorithm, matching information collected in Table~\ref{tab:perf2} and represented in Figure~\ref{density}, is the ANN (see Figure~\ref{den:f7}) for which the distribution of the predicted probability for INNs shows a peak close to one, although it maintains a second small peak close to zero. 
In order to further increase the performance, instead of using only one algorithm,  we consider a mixture of the two (BAG-ANN), in which the predicted probability is a convex linear combination of the predicted probabilities originated from the two algorithms independently\footnote{As a robustness check, we try the mixture of different algorithms, which do not lead to any improvement in the performance.}. The mixture weights are defined according to a function which maximises the separation between the predicted probabilities for innovative and non-innovative and the area under the ROC curve (AUC) (see Appendix~\ref{cutofs} for further details). Thereby, we construct a mixture of the two algorithms with weights 0.77 (BAG) and 0.23 (ANN) with the resulting predicted probability represented in Figure~\ref{fig:dens5}. Despite the overall improvement, there is still a large area of common support between INNs and NOINNs densities and this issue is particularly severe for intermediate values of the predicted probability. Prediction in that area inevitably leads to high type I and II errors, since there is not much difference between the two densities. The reason of this poor performance lay in the nature of the empirical problem. Classification algorithms perform well when the underlying nature of the variable to be classify is a categorical one. However often, and as it is in this case, the categories are the results of an artificial categorisation or dicothomisation of an otherwise continuous variable. The case of innovativeness is an exemplary one: firms can be more or less innovative on a continuum scale. For this reason, when using the model for prediction, instead of introducing only one cut-off which separates a predicted INN from a predicted NOINN, we identify two cut-offs which identify three intervals in the (0,1) domain of the predicted probability. Firms with a predicted probability smaller than the first cut-off are classified as NOINNs while firms with a predicted probability higher than the second cut-off are classified as INNs. We consider unclassified firms with a predicted probability in-between the two cut-offs and we will drop them from the analysis. In this specific case, we predict as NOINNs 2008 firms with a predicted probability smaller than 0.2 and as INNs those with a predicted probability higher than 0.8. The resulting confusion matrix is presented in Table~\ref{tab:perf_bag}. The algorithm turns out to be extremely performing in correctly classifying INNs: most of the misclassification errors are indeed false negative. This type of error reduce the differences among groups:  if we find a difference between innovative and non-innovative firms, the result would hold \textit{a fortiori} in a better algorithm.

\begin{table}[htbp!]
\caption{BAG-ANN mixture performances on the SMOTE 2013 start-up with different predicted probability cut-offs}\centering
\begin{footnotesize}
\begin{tabular}{c|cccccc}
  \hline\hline
\multirow{3}{*}{Real data} & \multicolumn{6}{c}{Prediction}                                                                            \\ \cline{2-7} 
                           & \multicolumn{2}{c|}{cut-off 0.2}     & \multicolumn{2}{c|}{cut-off 0.8}     & \multicolumn{2}{c}{Final subsample} \\
                           & NOINNs & \multicolumn{1}{c|}{INNs} & NOINNs & \multicolumn{1}{c|}{INNs} & NOINNs       & INNs       \\ \hline
NOINNs                   & 6,751     & \multicolumn{1}{c|}{2,274} & 8,698      & \multicolumn{1}{c|}{327}  & 6,751          & 327       \\
                           & 75\%     & \multicolumn{1}{c|}{}     & 96\%     & \multicolumn{1}{c|}{}     & 95\%           &            \\
INNs                       & 40       & \multicolumn{1}{c|}{110}  & 99       & \multicolumn{1}{c|}{51}   & 40            & 51        \\
                           &          & \multicolumn{1}{c|}{73\%} &          & \multicolumn{1}{c|}{34\%} &                & 56\%  \\
                           
                           \hline\hline

\end{tabular}
\end{footnotesize}
\label{tab:perf_bag}
\end{table}


\subsubsection{Predicting the past: innovative firms in 2008} \label{sec:pred}

The mixture model BAG-ANN, with  weights and cut-offs as discussed in Section~\ref{sec:tt}, can now be leveraged to predict which firms would have been innovative in the 2008 sample. In such a way, we are able to enrich the 2008 data with a new variable, namely $Inno$, which takes value 1 if the algorithm assign to a firm a predicted probability larger than 0.8 while value 0 if the algorithm assigns a predicted probability smaller than 0.2. We consider unclassified the other cases which, as Table~\ref{tab:perf_bag_08} shows, are about 10\% of the sample. This represents the learning step of our methodology.

\begin{table}[htbp!]
\caption{BAG-ANN mixture classification of NOINNs (predicted probability $\leq$ 0.2) and INNs (predicted probability $\geq$ 0.8) start-ups on the 2008 sample }\centering
\vspace{10pt}
\vspace{-0.5cm}
\begin{footnotesize}
\begin{tabular}{c|cc|c|cc|c}

  \hline  \hline
  &\multicolumn{2}{|c|}{Predicted Probability}&Total&\multicolumn{2}{|c|}{\%}&\%  \\
 & $\leq$0.2 & $\geq$ 0.8  &   & $\leq$0.2 &  $\geq$ 0.8& $\geq 0.2 \leq 0.8$ \\
  \hline
2008 &34,487   &763& 39,295 &87.8\% &1.9\% & 10.3 \%\\
 \hline\hline
\end{tabular}
\end{footnotesize}
\label{tab:perf_bag_08}
\end{table}

\noindent However, it is impossible to directly evaluate the performance of the 2008 prediction. It depends primarily on the assumption that the true and unknown model which generate the data in 2013 is similar to the one in 2008. 
Nevertheless, we can indirectly provide some statistics on the predicted INNs and NOINNs for a qualitative evaluation of the BAG-ANN performance. 
To this aim, Table~\ref{tab:comparison} shows the percentage of 2008 firms involved in R\&D and IPR investment and the average investment for the period 2008-2018 for INNs and NOINNs and the value are significantly higher for the former.

\begin{table}[ht]
\caption{Qualitative evaluation of the prediction}\label{tab:comparison}
\vspace{-0.3cm}
\centering
\begin{footnotesize}
\begin{tabular}{lrr}
  \hline
  \hline
 & INNs & NOINNs \\ 
  \hline
 \% of firms with positive R\&D investment over 10 years & 6\% & 4\% \\ 
  \% of firms with positive IPR investment over 10 years & 10\% & 4\% \\ 
   average R\&D investment over 10 years, if positive  & 612K & 346K \\ 
 average IPR investment over 10 years, if positive  & 7,056K & 7,76K \\
   \hline \hline
   \multicolumn{3}{r}{\footnotesize{Note: Differences between groups are statistically significant at the 1\%}}
\end{tabular}
\end{footnotesize}
\end{table}

%
\begin{table}[htbp!]
\caption{ATECO classification according to the BAG-ANN mixture classification of NOINNs and INNs on the 2008 sample}\centering
\begin{footnotesize}
\begin{tabular}{l|cc|c|cc}
  \hline\hline
 & predicted&probability& Total & ATECO \%& ATECO \% \\ 
 ATECO& $\leq$0.2 & $\geq$ 0.8&  & $\leq$ 0.2& $\geq$ 0.8\\ 
  \hline\hline
  A & 743 &   12 & 805   & 92.23\% &   1.49\% \\ 
  B &  41 &   1  &  44   & 93.18\% &  2.27\% \\ 
  C & 3,222 &  109& 3,903   & 82.55\% &  2.79\% \\ 
  D & 805 &  17 & 880    & 91.48\% &   1.93\% \\ 
  E &  145 &   1 &  170   & 85.29\% &   0.58\% \\ 
  F & 6,936 &  136& 7,869  & 88.14\% &  1.73\% \\ 
  G & 5,885 &  195& 7,014  & 93.90\% &  2.78\% \\ 
  H & 946 &   30 & 1,146   & 82.55\% &   2.62\% \\ 
  I & 1,721 &  56 & 2,013   & 85.49\% &   2.78\% \\ 
  J & 1,537 &  37 & 1,754   & 87.62\% &   2.11\% \\ 
  K & 575 &   7 & 627    & 91.71\% &   1.12\% \\ 
  L & 4,763 &  48& 5,088   & 93.36\% &  0.94\% \\ 
  M & 3,294 &  49& 3,646   & 90.35\% &  1.34\% \\ 
  N & 1,754 &  32 & 1,982   & 88.50\% &   3.85\% \\ 
  O &   2 &   0 &   5   & 40.00\% &   0.00\% \\ 
  P & 372 &   5 & 401    & 92.77\% &   1.25\% \\ 
  Q & 594 &   8 & 681  & 87.22\% &   1.17\% \\ 
  R & 689 &   13 & 762   & 90.42\% &   1.71\% \\ 
  S & 423 &   6 & 467    & 90.58\% &   1.28\% \\ 
  NA & 37 &   1  &   38 & 97.37\% & 2.63\% \\
   \hline
  Total &   34,487 &763  & 39,295   &&   \\ 
   \hline\hline
\end{tabular}
\end{footnotesize}
\label{tab:08_ebiemp2}
\end{table}

\section{Econometric analysis}\label{analysis}
In this session we test the hypothesis of a survival premium of INNs, that is 2008 start-up classified as innovative with respect to the NOINNs.

\paragraph{Univariate analysis}
We first employ the Kaplan-Meier estimator (KME) to show short- and long-term differences, within 2008 firms, on their ability to survive in the Italian market during the crisis. 

The KME is a non-parametric estimator classically used, among the others, to estimate  survival distribution functions \citep[see][for a discussion on its statistical properties]{Fleming, Andersen}.
In general, this analysis studies the time to death for a population with survival distribution function $S(t)$, namely the probability that a start-up will be still alive at time $t$. Let consider a sample from the population with dimension $n$ (note that here we are dealing with a right-censoring problem). 
Denote with $t_1 < t_2 < \cdots$ the years when start-ups definitely close their actives on the Italian market. Let $d_i$ be the number of start-ups who close at $t_i$. The Kaplan–Meier estimator $\hat{S}(t)$ for $S(t)$ is:
\begin{equation}
    \hat{S}(t)=\prod_{t_i\leq t}\left(1-\frac{d_i}{r_i}\right)
\end{equation}
where $r_i$ is the number of start-ups in the risk set just before time $t_i$, \textit{i.e.} that firms who had survived, and $d_i$ the number of failures at time $t_i$.  The variance of the KME is estimated by the Greenwood’s formula:
\begin{equation}\label{eq:1}
    \hat{\sigma}^2(t)=\hat{S}^2(t)\sum_{t_i\leq t}\left(\frac{d_i}{r_i(r_i-d_i)}\right).
\end{equation}
Eq.~\eqref{eq:1} represents the standard error assigned to the KME using the Delta-method. Note that, for large samples, the KME is approximately normally distributed so the marginal error at $(1-\alpha)$ confidence level is  $z_{1-\alpha/2}\hat{\sigma}(t)$.
Finally, the confidence interval, also for quite small sample sizes \citep{borgan1990note}, taking advantage of the log-minus-log transformation, is 
\begin{equation}
    \hat{S}(t)\exp\left\{\pm \frac{z_{1-\alpha/2}\hat{\sigma}(t)}{\hat{S}(t)\ln{\hat{S}(t)}}\right\}.
\end{equation}

\begin{figure}[htbp!]
\begin{center}
\centerline{\includegraphics[scale=0.45]{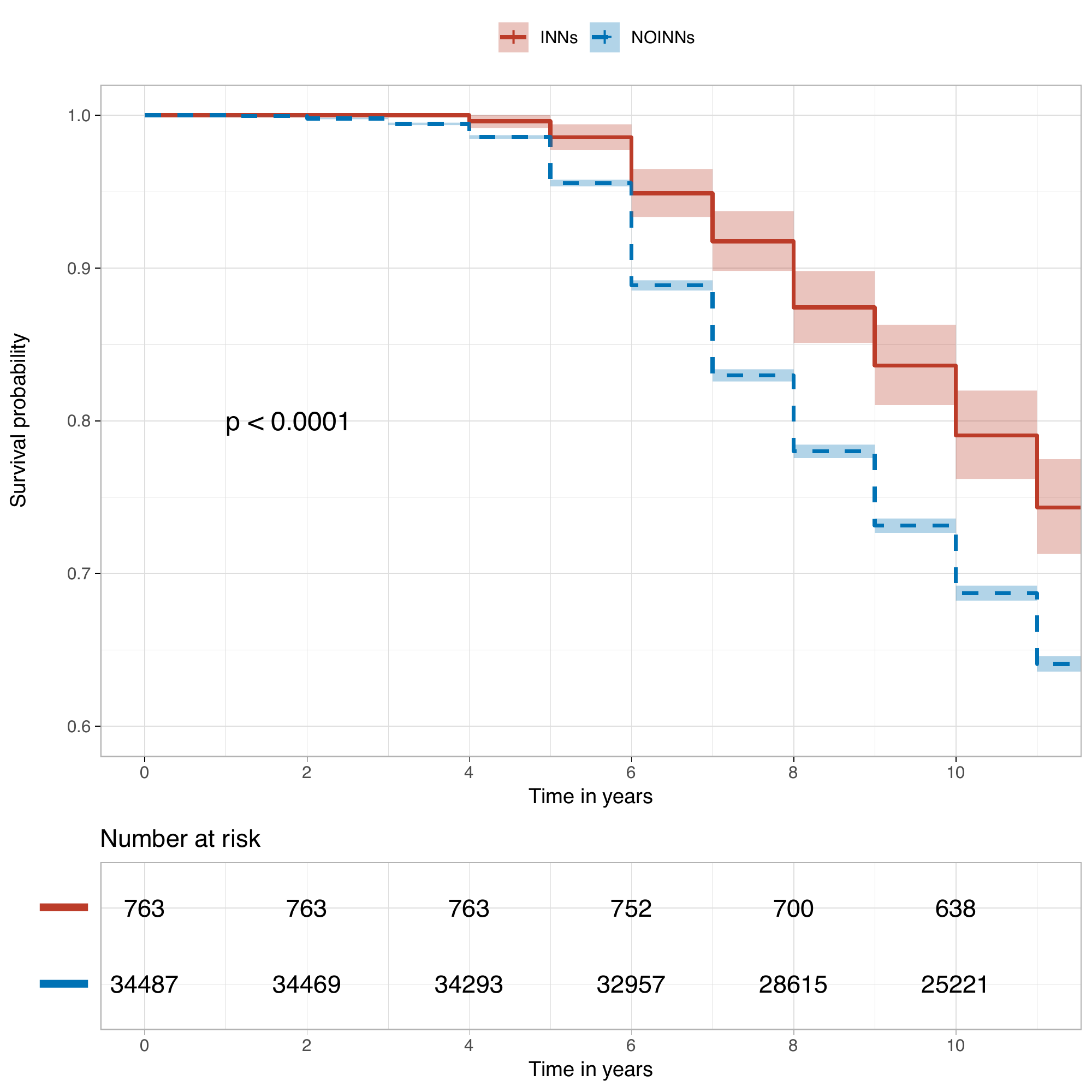}} \caption{Survival curves of INNs and NOINNs. P-value for log-rank test is reported. Log-minus-log transformation is applied for confidence intervals}\label{fig:surv2}
\end{center}
\end{figure}

\noindent The KME allows for direct comparisons across the survival probability of samples with different sizes. Figure~\ref{fig:surv2} shows the two Kaplan-Meier curves
with time in years on the horizontal axis and probability of surviving, or proportion of firm surviving, on the vertical one. Lines represent the survival curves stratified by INNs and NOINNs within their shadows of confidence intervals. At time zero, the survival probability is 1.0 (namely 100\% of the firms are alive). After ten years, the survival probability is approximately 0.687 (or 68.7\% - standard deviation 0.002497) for NOINNs and 0.790 (or 79.0\% - standard deviation 0.01474) for INNs. INNs enjoy a survival premium and their survival curve lays always above NOINNs. The associated confidence intervals are wider for INNs than NOINNs, suggesting the higher uncertainty around innovative ventures. Nevertheless, there is always a statistically significant difference in the two groups as the rejection of the null-hypothesis of the log-rank suggests.
  
\paragraph{Multivariate analysis} 
We now seek to address the last issue raised in the theory, that is whether the survival premium of innovative firms persists even when controlling for sectors and locations. We perform this task  by adding the one-digit ATECO2007 classification for economic activities, the NUTS3 region classification (namely ``provincia'') for the location effect, and the interaction variables of being INNs with both sector and region as controls in a Cox proportional hazards model. The interaction effect can be interpreted as the positive or negative survival premium linked with the specific sector and region. With the Cox model in Eq~\eqref{eq:cox}, we simultaneously estimate the impact of several variables on the survival. More precisely, we estimate how the effect of being INN in a specific sector and in a given location influences the exit rate from the market at a particular year, given that a firm survived up to that year. \emph{Id est}, the hazard rate of failure at time $t$ is

\begin{equation}\label{eq:cox}
    h(t)=h_0(t)\, \exp\left\{\beta_1*Inno+\beta_2*Sector +\beta_3*Location+\beta_4*Interactions\right\}
\end{equation}

\noindent where:
\begin{itemize}
    \vspace{-0.3cm}\item $t$ is the survival time;
\vspace{-0.3cm}\item $h(t)$ is the hazard function;
\vspace{-0.3cm}\item $\beta_i$ are the coefficients. Since the Cox model can be written as a linear regression model of the logarithm of the hazard, it is possible to interpret the $exp(\beta_i)$ as the hazard ratio of the $i^{th}$ covariate;
\vspace{-0.3cm}\item $h_0$ is the baseline hazard when all the covariates are set equal to zero.
It is possible to estimate the $\beta_i$ without any consideration of the hazard function only under the assumption of proportional hazard, validated both visually and with the log-rank test (see Table~\ref{results};
\vspace{-0.3cm}
\item $Inno$, $Sector$ and $Location$ are categorical variables summarised in Table~\ref{tab:vars}, while $Interaction$ are the interaction terms between $Inno$ and the remaining variables.
\end{itemize}

\begin{table}[!htbp]
\centering
\caption{Variables description}
\label{tab:vars}
\begin{footnotesize}
\begin{tabular}{l|l|c|l|c}
\hline
\hline
\textbf{Name} & \textbf{Description} & \textbf{Categories} & \textbf{Reference} & \textbf{Observation} \\ \hline
$Inno$ & \begin{tabular}[c]{@{}l@{}}Dummy variable for being \\ an INN or a NOINN \end{tabular} & 2 & NOINNs & 35,250 \\
\hline
$Sector$ & ATECO classification of sectors & 20 & Manufacturing & 35,212 \\
\hline
$Location$ & \begin{tabular}[c]{@{}l@{}}Italian Province (NUTS3 region) in\\ which firm is located\end{tabular} & 110 & Milan & 35,250\\
\hline
\hline
\end{tabular}
\end{footnotesize}
\end{table}

\begin{figure}[!tbp] 

  \centering
  \subfloat[NUTS3 regions]{\includegraphics[width=0.5\textwidth]{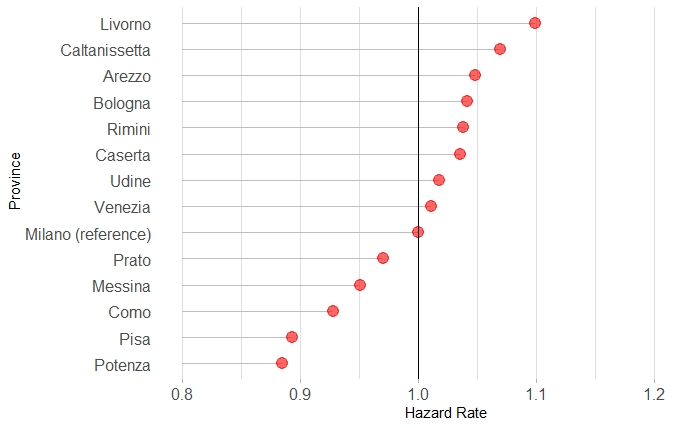}\label{fig:f1}}
  \hfill
  \subfloat[ATECO sector]{\includegraphics[width=0.5\textwidth]{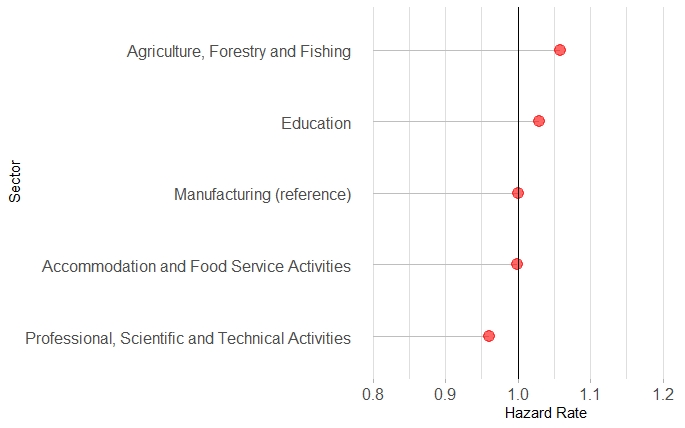}\label{fig:f2}}
  \caption{Hazard Ratio for the interaction term}
  \label{fig:inter}
\end{figure}

\noindent Table~\ref{results} summarises the results for five different models and estimated coefficients. Model (1) uses just the dummy variable for INNs. The coefficient value -0.428 shows that being innovative has a negative and statistically significant effect on the probability of failure with respect to NOINNs. A straight interpretation of the effect is to compute the hazard ratio $=e^{-0.428}=0.65$. \emph{I.e.} at any given time, innovative firms almost double their chance of survival vis-\`a-vis NOINNs. Models~(2) and~(3) add industrial sector and regional controls, respectively, while models~(4) and~(5) consider also their interaction effects with INNs. When adding interaction effects for the location, the significance of being innovative fades. This evidence suggests that, as pointed by the theoretical consideration \citep{feldman2001entrepreneurial}, a large part of the survival premium experienced by INNs depends on a self-selection of innovative firms for locations in which any firm, and not only innovative ones,  is more likely to survive.
We like to make few considerations. This result does not imply that being innovative is irrelevant. For instance, being innovative in a specific region can still lead to a survival premium. By looking at the interaction of location with the innovative dummy, we can rank Italian province according to the survival premium for being innovative. Figure~\ref{fig:inter} shows the hazard ratio of the interaction effects when statistically significant. The higher the value, the higher is the positive effect of innovation on the chance of survival.
Second, it might seem counter-intuitive that sector controls do not absorb the explanation power of INNs, whereas location does. NUTS3 regions can capture a much larger effect which includes, on the one hand, the mix of sectors characterising a geographical areas and, on the other hand, dynamics discussed above such as entrepreneurial atmosphere, agglomeration economies, university roles, and so on. However, also for sectors, we can compute the magnitude of the interaction effect as plotted in Figure~\ref{fig:inter}. An inquiry on the causes that make INNs more likely to survive in some locations or sectors is outside the scope of this work, but it surely leave room for new research questions. Note that, at least theoretically, a further model based on the joint estimate of both sectors and locations is possible. Unfortunately, here, especially for the 2008 firms classified as INNs, we suffer from the complete separation problem, which does not allow the estimation of some interaction effects.

\begin{table}[!htbp] \centering \footnotesize
  \caption{Cox regressions: summary} 
  \label{results} 
\begin{tabular}{@{\extracolsep{5pt}}lccccc} 
\\[-1.8ex]\hline 
\hline \\[-1.8ex] 
 & \multicolumn{5}{c}{\textit{Dependent variable:}} \\ 
\cline{2-6} 
\\[-1.8ex] & \multicolumn{5}{c}{Survival} \\ 
\\[-1.8ex] & (1) & (2) & (3) & (4) & (5)\\ 
\hline \\[-1.8ex] 
$Inno$ & $-$0.428$^{***}$ & $-$0.459$^{***}$ & $-$0.438$^{***}$ & $-$0.512$^{***}$ & $-$0.122 \\ 
  & (0.072) & (0.072) & (0.072) & (0.198) & (0.246) \\ 
  & & & & & \\ 
$Sector$ &   & YES &  & YES &  \\ 
$Location$ &  &  &YES  &  &YES  \\ 
$Inno$ * $Sector$ &  &  &  & YES &  \\ 
$Inno$ * $Location$ &  &  &  &  &  YES \\ 
 
\hline \\[-1.8ex] 
Observations & 35,250 & 35,212 & 35,250 & 35,212 & 35,250 \\ 
Log Likelihood & $-$129,172.400 & $-$128,598.200 & $-$128,991.200 & $-$128,591.100 & $-$128,940.600 \\ 
Wald Test & 35.340$^{***}$ & 483.320$^{***}$  & 381.580$^{***}$  & 490.800$^{***}$  & 386.730$^{***}$  \\ 
LR Test & 40.756$^{***}$  & 493.846$^{***}$  & 403.239$^{***}$  & 508.001$^{***}$  & 504.494$^{***}$  \\ 
Score (Logrank) Test & 35.885$^{***}$  & 493.376$^{***}$  & 388.083$^{***}$  & 504.257$^{***}$  & 443.604$^{***}$  \\ 
Df  & 1 & 19 & 110 & 36 & 209 \\ 
\hline 
\hline \\[-1.8ex] 
\textit{Note:}  & \multicolumn{5}{r}{$^{*}$p$<$0.1; $^{**}$p$<$0.05; $^{***}$p$<$0.01} \\ 
\end{tabular} 
\end{table}

\section{Conclusion}\label{sec:conclu}

This paper contributes to design a new research framework combining data science within econometric models. In particular, we use machine learning algorithms to extrapolate information from large source of data, which could have not been otherwise employed in standard regression models. We stress and show how this exercise needs feedback between theory, econometrics and data science, to design the desirable properties of the variable created by machine learning algorithms. 

We apply this methodological approach to a long standing debate in economics of innovation and, specifically, we develop a new indicator of innovation at firm level which removes some drawbacks of previous proxies.
Beginning with the precise theoretical problem of assessing the survival premium of innovative start-ups, we employed supervised machine learning techniques to create the indicator. In this process, we carefully considered specific constraints to make the new variable suitable for the use in an econometric model. The machine learning algorithm is trained on all possible information, except variables on location and sector as well as on R\&D and IPR. In this way, it has been possible to run a survival model which, as suggested by the theory, includes sector and location as controls and does not suffer from any form of endogeneity. 
 
 This framework can be considered as a mild integration between econometrics and data science since the two approaches are connected via feedback, but they still run separately. Also, it is possible to imagine different scenarios in which the methodological integration is higher or in which data science completely supersedes econometrics. However, given the state of art of the economic science, which stresses very much the importance of causal relationships, we believe that these research frameworks are yet to be designed.

As a second contribution, we introduce a new indicator for innovation which is a suitable candidate to be used in various analysis since it overcomes  many of the main drawbacks of other innovation proxies: it blends together different aspects of both inputs and outputs of the innovation process. However, its nature is very much connected with the Italian case. In this paper, we use the model to predict the innovativeness of Italian start-ups in the past, but the same exercise can be done to predict the innovativeness of foreign firms in the present. Indeed, the AIDA Bureau van Dijk database used to train and test the algorithm is consistent with the ORBIS-AMADEUS database which collects the same information as AIDA on European firms. Nevertheless, the application of a prediction algorithm on sample in a different country requires considerable efforts in evaluating the results in relation with other measures of innovation such as patent applications or survey data. However, for the largest European economies, a match between AMADEUS-ORBIS with PATSTAT and CIS data is already in place. Thus, a major line for future work opened up by this paper is the extension of this new measurement to other countries.

On the basis of the new indicator, as a third contribution, we provide new empirical evidence on survival of INNs. When controlling for sector specificity, INNs seem to maintain their survival premium, which conversely fades out when controlling for the location at the NUTS3 regional level. This result challenges previous literature which formed a weak consensus on a positive effect of innovativeness on survival. We find that INNs have a survival premium only in relation with specific locations. Probably, the specific attributes  of a location, which include also the composition of the local economy in term of sectors, might be more or less suitable for a newly established innovative firm. For many locations in the dataset, the effect is not statistically significant while for others it could not even be estimated due to the small number of start-ups in that areas; however, for some regions a clear-cut effect exists. Understanding the determinants of survival at the regional level could be a question to be addressed in further work.

\clearpage

\bibliography{biblio}

\newpage
\appendix
\section{Appendix}

\subsection{Missing Value analysis} \label{mva}
The AIDA database is a valuable source of information but only the collection of some variables is mandatory. Hence, a missing value analysis (MVA) is needed to avoid information loss when applying  machine learning algorithms, which immediately discard all observations containing missing values (NAs). We propose a MVA to identify variables and observations containing the highest NA amount and to choose which ones to delete. It is a semi-automatic approach which balances information loss and introduction of source of extra variability. No imputation of missing data is undertaken to avoid introducing potential bias in variables with too many NA\footnote{Note that NAs are too much diffused among variables and observations, therefore multiple imputation will add an extra variability to observed variables not justified. 
Even limiting the multiple imputation to some crucial variables, we do not have enough complete observations in the dataset to finalise the NA completion.}.

\begin{figure}[htbp!]
\begin{center}
\includegraphics[scale=0.39]{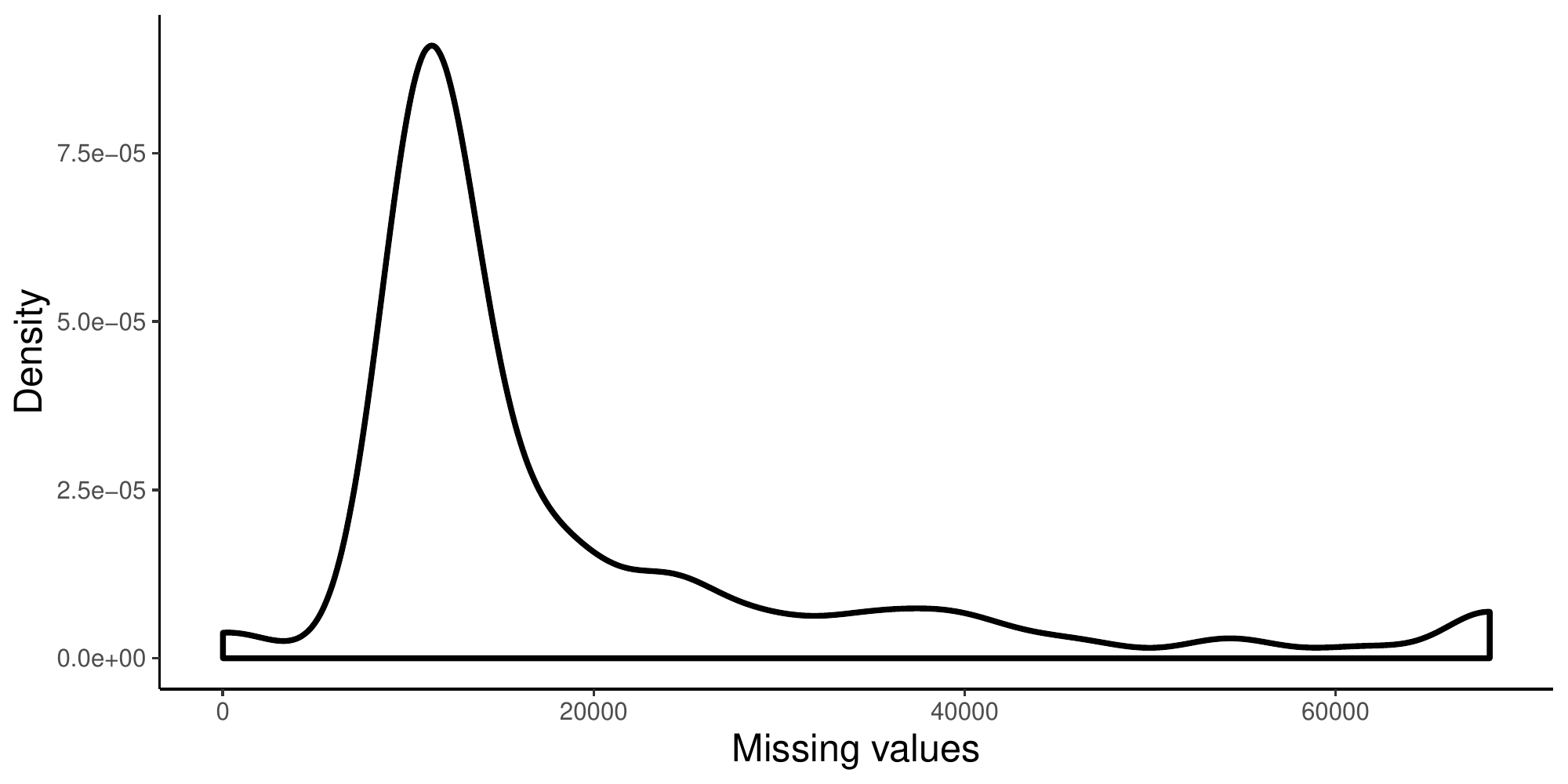}
\includegraphics[scale=0.39]{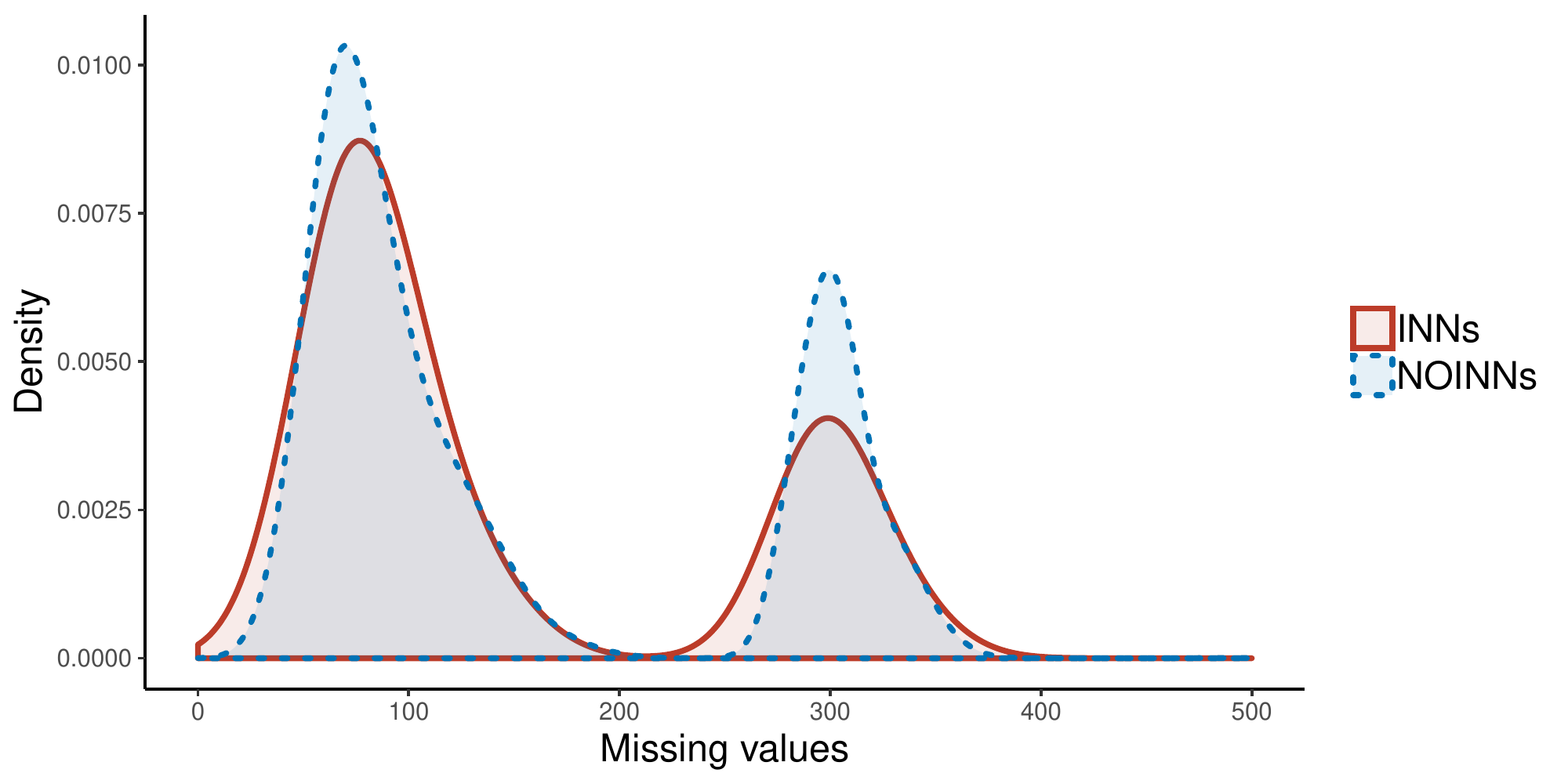}
\caption{The left panel shows the missing value distribution  in 538 variables for the 68,316 observed 2013 start-ups. The right panel proposes missing value distributions in observations separated in INNs ($n=1,010$) and NOINNs ($n=67,306$), over 538 variables}\label{fig:na}
\end{center}
\end{figure}

The MVA starts with the 2013 sample and only afterwards evaluates the status of the 2008 one.
Since in the 2013 sample variables measured in 2015 had not yet been fully incorporated into AIDA at the time of the inspection (July 2016), we chose to drop them directly\footnote{Note that management variables, which contain a huge amount of not uniformed  text, are discarded since the beginning of the data construction process.}. We start with 800 (174 do not change over time while 786 are the results of the firm observation over three years, namely 262 $\times$ 3) available variables and we discard 262 accounting variables, i.e. we still retain 538 for the 68,316 start-ups of 2013. We also control for the presence of duplicated variables. Subsequently, we define the number of NAs for each variable and for each firm, obtaining the distributions proposed in Table~\ref{tab:na} and Figures~\ref{fig:na}. 
We observe that the NAs affect in a similar way both INNs and NOINNs. 
We choose to drop observations with a number of NAs higher than the third quartile, i.e. with more than 290 missing over the 538 variables. We obtain a data-set composed by 51,496 observations (including 796 innovative start-ups). 

Then, we drop variables with more than 3,968 (equals to the first quartile) NAs. We retain 51,496 firms observed on 127 variables. 
We undertake the same analysis also on the 2008 sample so as not to loose too many firms from the 2008 sample. After removing variables already discard in the 2013 sample, three new variables containing more than the 30\% of missing values are identified. Hence, we drop them from both samples and we discard all observations  still containing NAs. We obtain two final datasets with 124 variables: the 2013 one retains 45,576 firms while the the 2008 one contains 
39,295 firms\footnote{Note that without doing this last MVA step in the 2008 sample only 18,078 firms would be left, representing less than the 28\% of the initial amount of 2008 start-ups.}.  
The 2008 sample is, then, enriched with further economic variables, such as EBITDA, R\&D investments, Employees, IPR investments observed from 2009 to 2018.
After the MVA, the proportion of INNs/NOINNs in the 2013 sample is consistent with the original one, slightly growing from 1.5\% to 1.59\%. 
\begin{table}[htbp]
\caption{Missing value distribution observed in 538 variables for the 68,316 observed 2013 start-ups according to INNs and NOINNs classification}\centering
\begin{footnotesize}
\begin{tabular}{llrrrrrrr}
  \hline  \hline
  Missing value&INNs/NOINNs&Min.  &1st Qu.   &Median     &Mean  &3rd Qu.     &Max.      \\ 
  \hline
Variables&INNs+NOINNs& 0   &11,170   &11,830   &19,530   &23,620   &68,320 \\
&INNs&0.00   &88.25  &213.00  &274.90  &297.00 &1010.00\\
&NOINNs&0   &11,050   &11,620  & 19,260   &23,370   &67,310\\
\hline
Firms&INNs+NOINNs&   24.0    &70.0    &98.0   &153.8   &290.0   &530.0 \\
&INNs&32.0    &71.0    &92.0   &146.4   &284.0   &360.0\\
&NOINNs   &24.0    &70.0    &98.0   &153.9   &290.0  & 530.0 \\
\hline\hline
\end{tabular}
\end{footnotesize}
\label{tab:na}
\end{table}

\begin{table}[htbp!]
\begin{center}
\begin{footnotesize}
\caption{Starting and final sample dimension description}
\begin{tabular}{lll}
\hline\hline
Sample       & 2008  & 2013  \\ \hline
Total        & 65,088 & 68,316 \\
NA           & 25,793 & 22,740 \\
Final sample & 39,295 & 45,576 \\ \hline\hline
\end{tabular}
\end{footnotesize}
\end{center}
\end{table}

\subsection{Algorithms}\label{algo}
For the train, test and learning step, we select seven algorithms known in machine learning, neural network and econometric literature, briefly presented in what follows. 

\begin{enumerate}
\item The binary recursive partitioning  algorithm (RPART)\citep{BFO84} is a tree-based method \citep{HTF08} grounded on a top-down approach in which the partition starts at the top of the tree. Starting from all observations in a single region,  the algorithm only successively splits the  space via a two further branches in the tree. Gini's coefficient is used for the tree variable selection.  RPART defines the best split at each step and predictions are easily interpretable, differently from other classification algorithms \citep{GWHT13}. We use RPART in $\mathsf{R}$ with the function \verb+rpart()+ in the package \verb+rpart+.

\item The classification tree (TREE) \citep{BFO84, Ripley} is based on binary recursive partitioning given the classification INNs/NOINNs. It recursively chooses splits from the selected independent variables. Numerical  variables are split at a given value $\alpha$ in each node, while  categorical variables are split accordingly to  two non-empty sets of unordered levels. At each step TREE selects the split which minimize classification impurity.
We use TREE in $\mathsf{R}$ with the function \verb+tree()+ in the package \verb+tree+.

\item The conditional inference tree (CTREE) \citep{HHZ06, SW99} estimates a regression relationship by binary recursive partitioning in a conditional inference framework. It uses a permutation test in order to select the set of variables that maximizes the Gini coefficient, differently from other tree based method that just select at each step one variable. We use CTREE in $\mathsf{R}$ with the function \verb+ctree()+ in the package \verb+party+.

\item 
The bagging algorithm (BAG) \citep{BFO84, SMT09, GWHT13}, or bootstrap aggregation, is based on the necessity to reduce the variance of the statistical learning tree previously described. It is simply based on the idea that the variance can be reduced if instead of only one training set, we use the average of more training sets. For this reason, it is based on the aggregation of many decision trees. BAG  generates $M$ different bootstrapped training data sets (with an increment in computation time), then it trains the method on the $M$ bootstrapped sets in order to average all the obtained predictions. Here we use RPART as the basis of BAG. We use BAG in $\mathsf{R}$ with the function \verb+bagging()+ in the package \verb+ipred+.

\item The Logit regression model (LOGIT) \citep{McCullagh}  is here used as the benchmark and widely used econometric model. It can be seen as a generalised linear model based on a Logit link function \citep{Agresti, McCullagh} or a random utility model for discrete choices \citep{Train}. 
It estimates a linear relation between the independent variables and the logit of the INNs probability. Its accuracy suffers in the presence of huge datasets, as in our case \citep{Perlich}.  We use LOGIT in $\mathsf{R}$ with the function \verb+glm()+.

\item The na\"ive Bayesian classifier (NB) \citep{Alpaydin}, in its particular binary version, 
is based on the estimation of the conditional a-posterior probabilities of INNs given the selected independent predictors, using the well-known Bayes rule. We use NB in $\mathsf{R}$ with the function \verb+naiveBayes()+ in the package \verb+e1071+. 

\item The artificial neural network (ANN)\citep{Bishop, Ripley} is a single hidden layer back-propagation network \citep{HTF08}. It is based on the artificial reproduction of the functioning of the brain \citep{Posner}, therefore ANN is a nonlinear statistical models based on a two-stage estimator. We use ANN in $\mathsf{R}$ with the function \verb+nnet()+ in the package \verb+nnet+. 
\end{enumerate}

\subsection{Optimal cut-off and mixture weight optimisation} \label{cutofs}
Part of the methodology introduced in this contribution is new, therefore new $\mathsf{R}$ functions has been coded to undertake the analysis. 
A first build-in function implements three criteria for the selection of the optimal cut-off in each algorithm:

\begin{enumerate}
    \item the Youden index (J) method which defines the optimal cut-point as the point maximising the difference between true positive rate and false positive rate (namely, the Youden function) over all possible cut-point values;
    \item the point closest-to-(0,1) corner in the ROC plane method which defines the optimal cut-point as the point minimising the Euclidean distance between the ROC curve and the (0,1) point;
    \item the optimal cut-point method which select as optimal cut-off the point maximising the product of sensitivity and specificity.
\end{enumerate}

\noindent The confusion matrix in Table~\ref{tab:perf2} is the result of the application of the second criterion. Similar results have been obtain applying the other two approaches. 

A second build-in $\mathsf{R}$ function finds the optimal mixture weights, following the approach below. First, we select two (or more) candidate algorithms to compose the mixture (here $alg1=$BAG and $alg2=$ANN), according to their performance emerging from the study of the ROC curve and of the confusion matrix.  Second, we retain the predicted probabilities ($pred.prod$), under the selected algorithms, for INNs (positive) and NOINNs (negative). Third, we select mixture weights $\alpha$ and $1-\alpha$ in the support (0,1) according to an optimisation process. The latter simultaneously maximises, for all $\alpha$ in the support, i) the Euclidean distance between INNs and NOINNs predicted probabilities and ii) the area under the ROC (AUC). A unique solution of this maximisation process exists and selects $\alpha$, such that the  predicted probability of the mixture is defines as follows: \[pred.prob.mixture=\alpha*pred.prod.alg1+(1-\alpha)*pred.prod.alg2.\]  

\subsection{Further descriptive statistics on the data}\label{app:descr}
\begin{figure}[htbp]
\begin{center}
\includegraphics[scale=0.39]{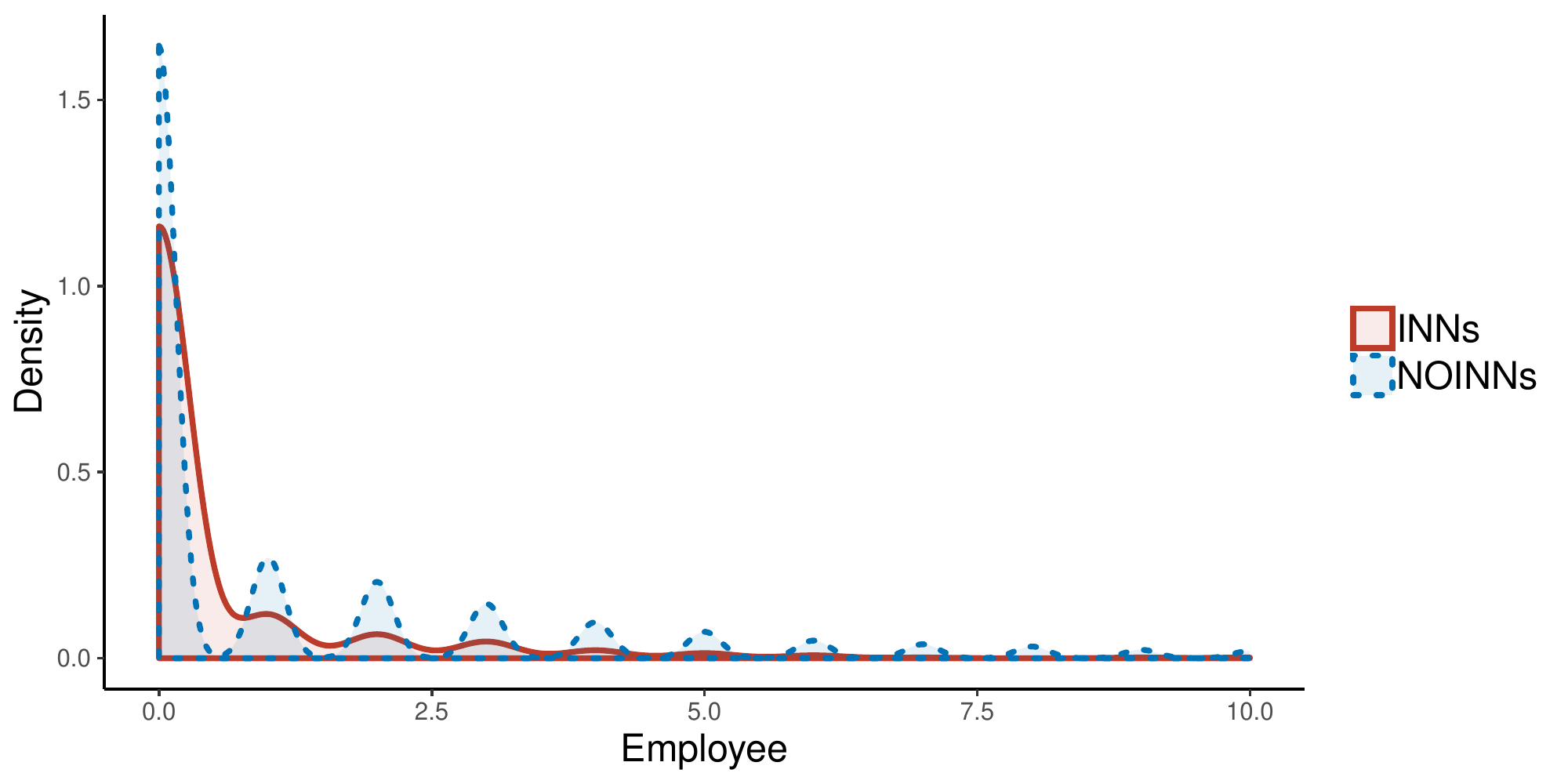}
\includegraphics[scale=0.39]{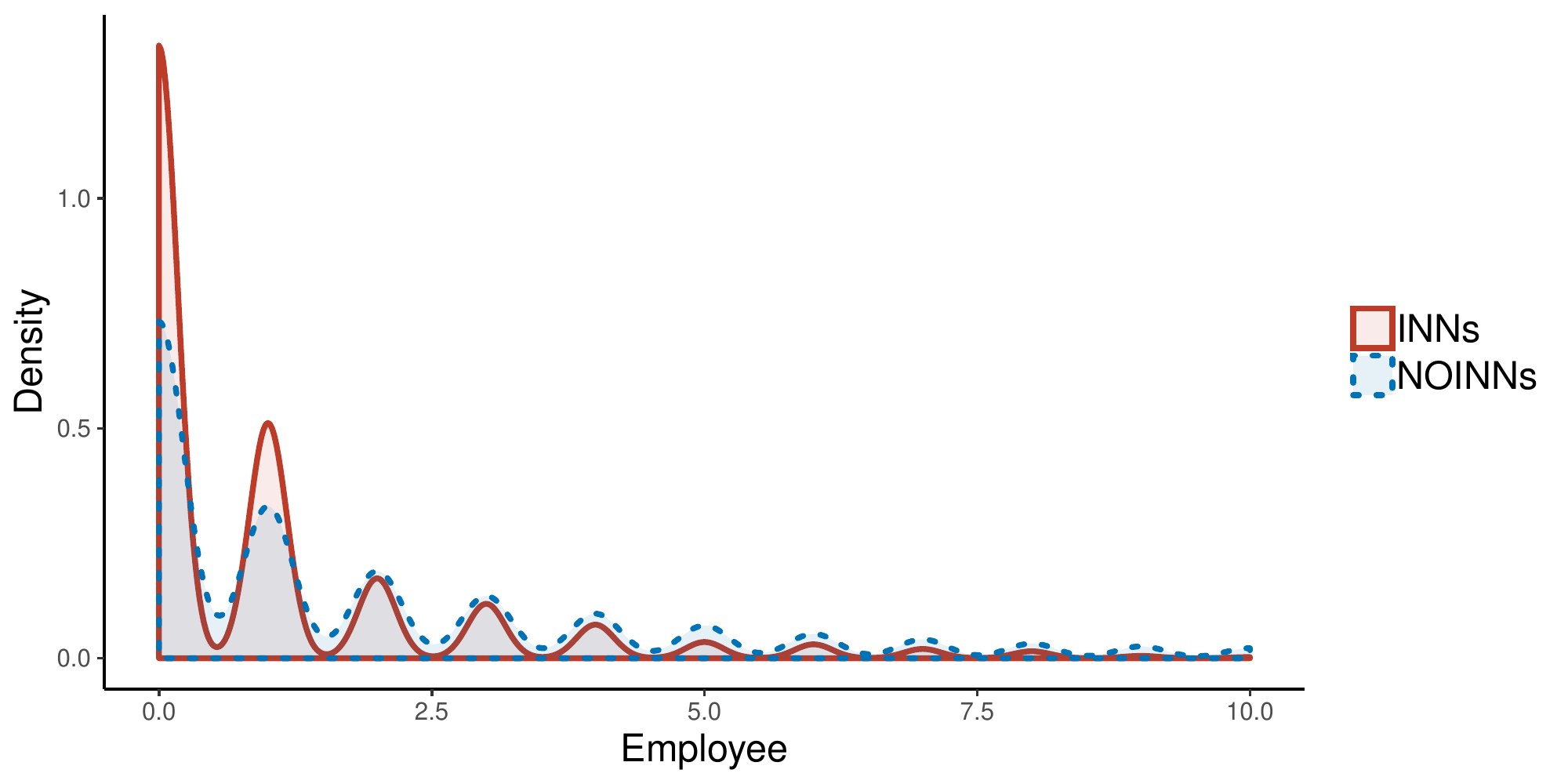}
\caption{Employees distributions in 2013 and 2014 of INNs and NOINNs in the 2013 sample}\label{fig:0}
\end{center}
\end{figure}

\begin{figure}[htbp]
\begin{center}
\includegraphics[scale=0.39]{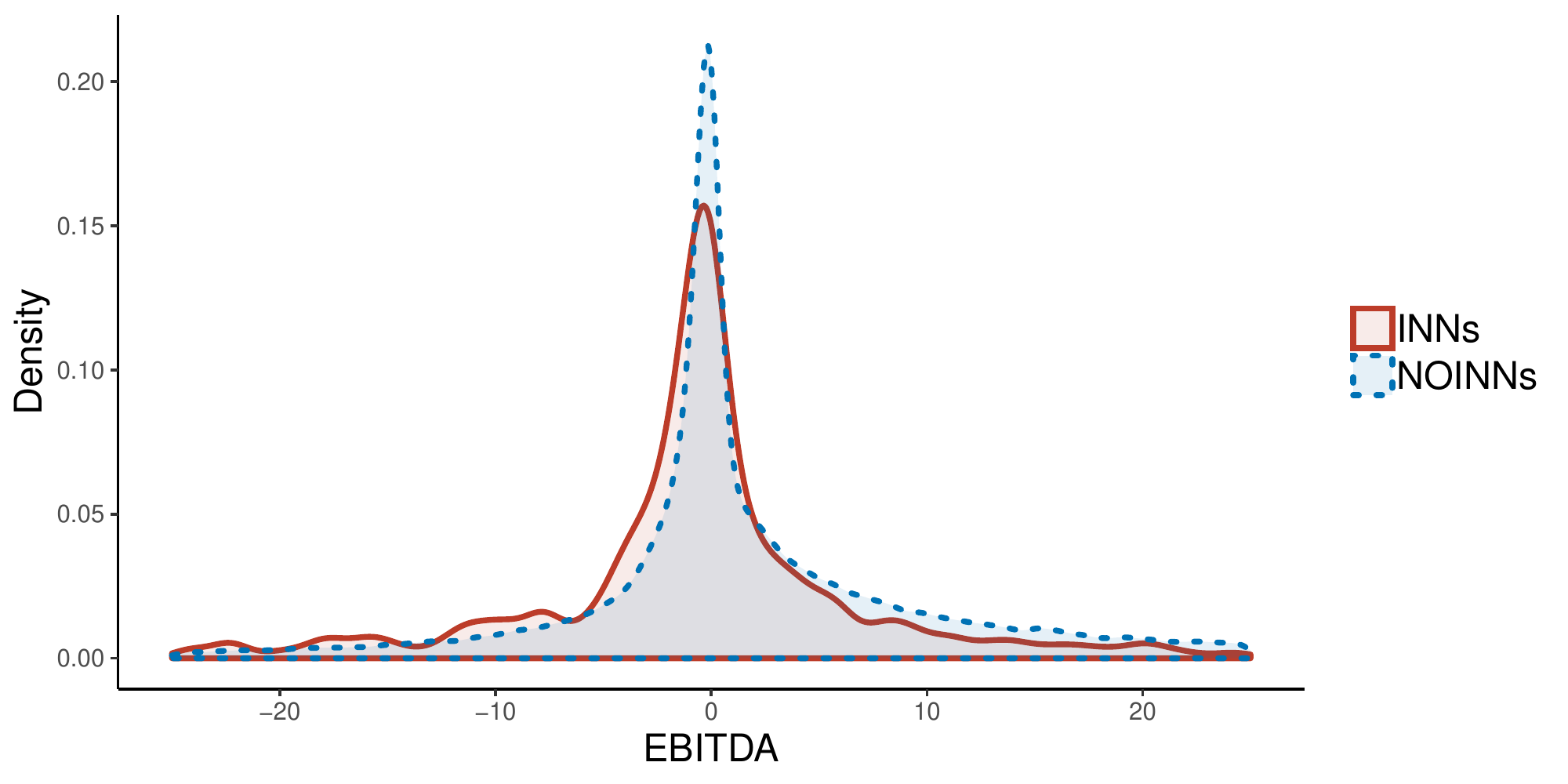}
\includegraphics[scale=0.39]{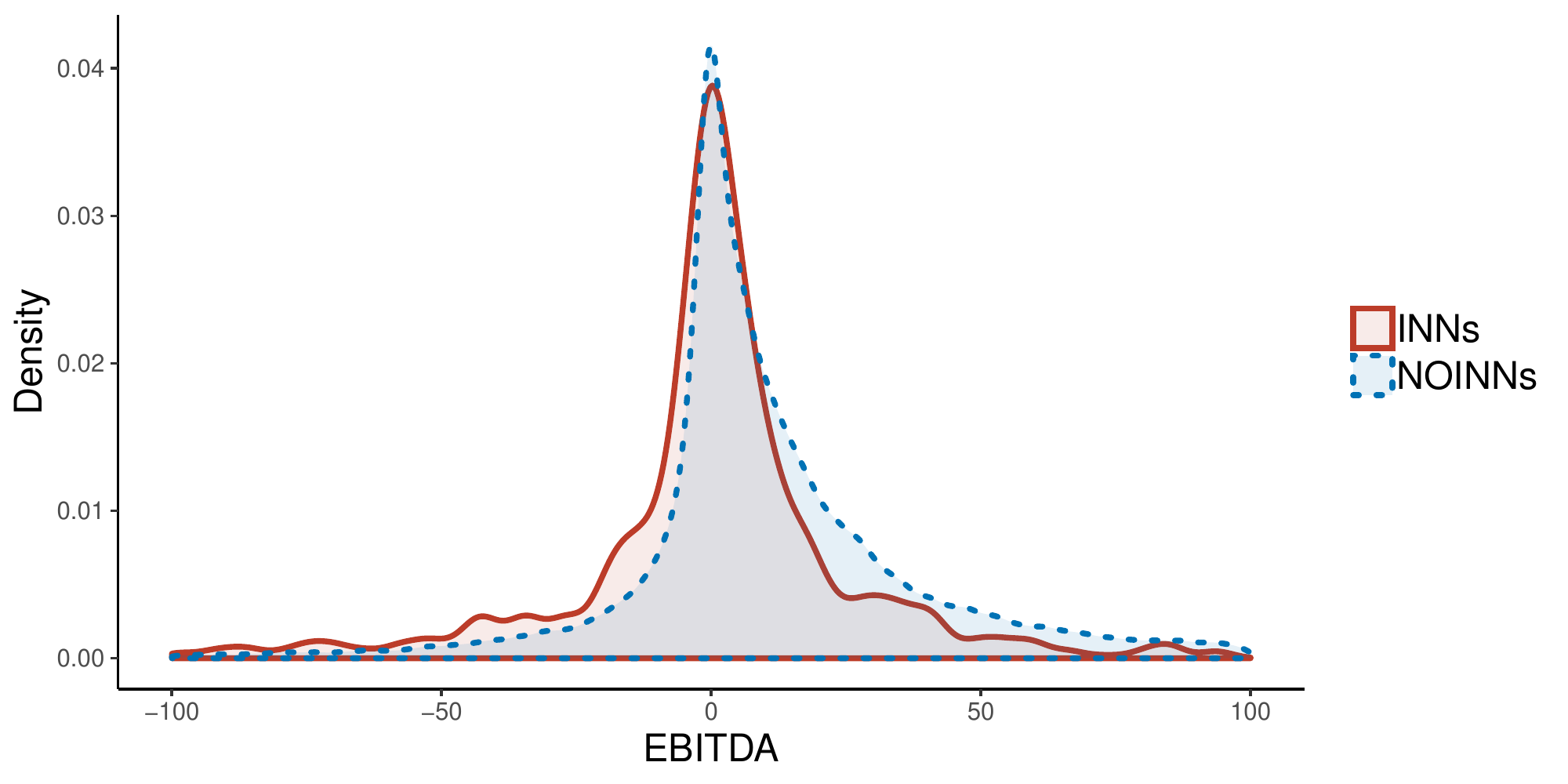} \caption{EBITDA distributions in 2013 and 2014 of INNs and NOINNs in the 2013 sample}\label{fig:0ebi}
\end{center}
\end{figure}

\begin{figure}[htpb]
\begin{center}
\centerline{\includegraphics[scale=0.45]{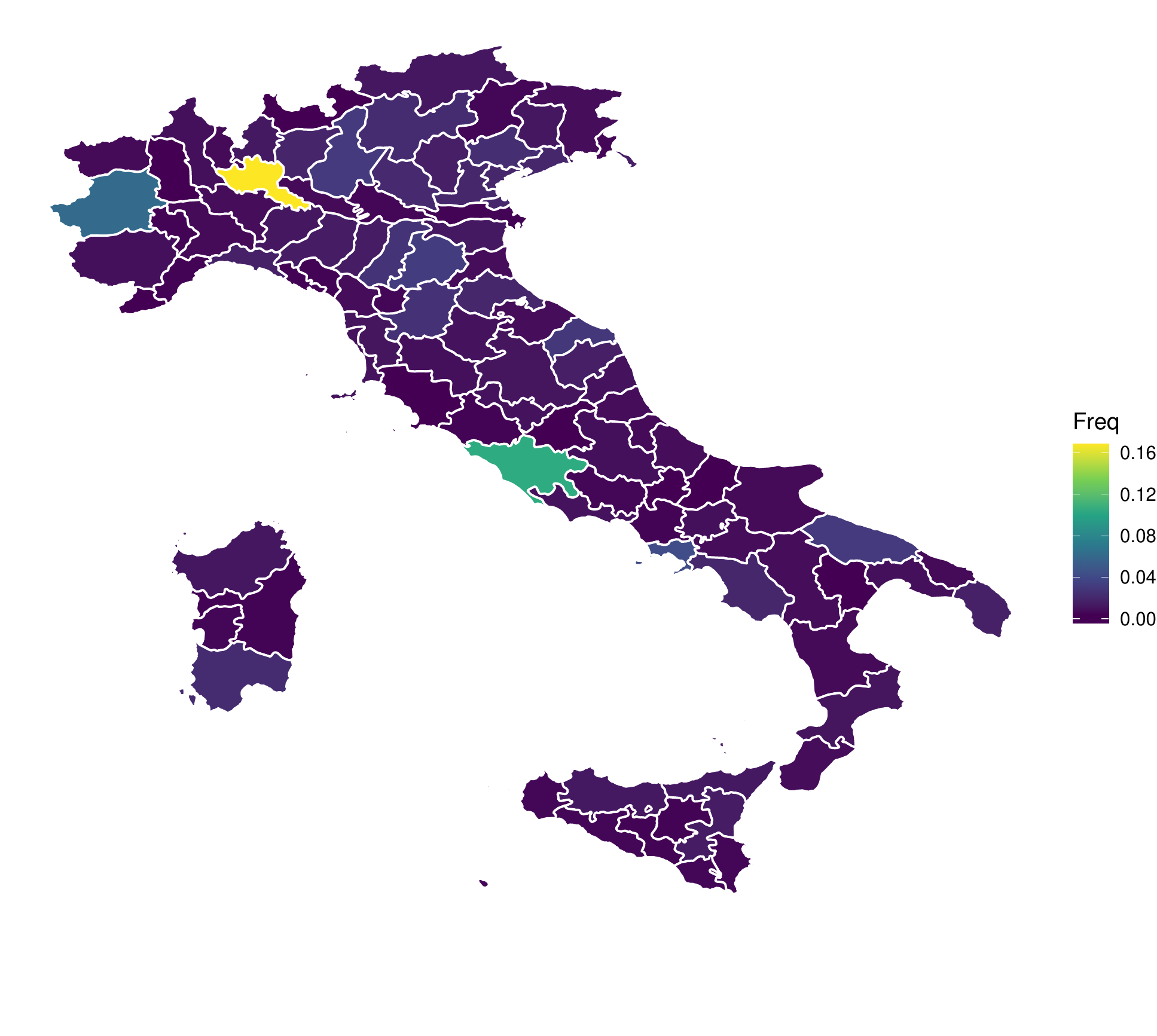}
\includegraphics[scale=0.45]{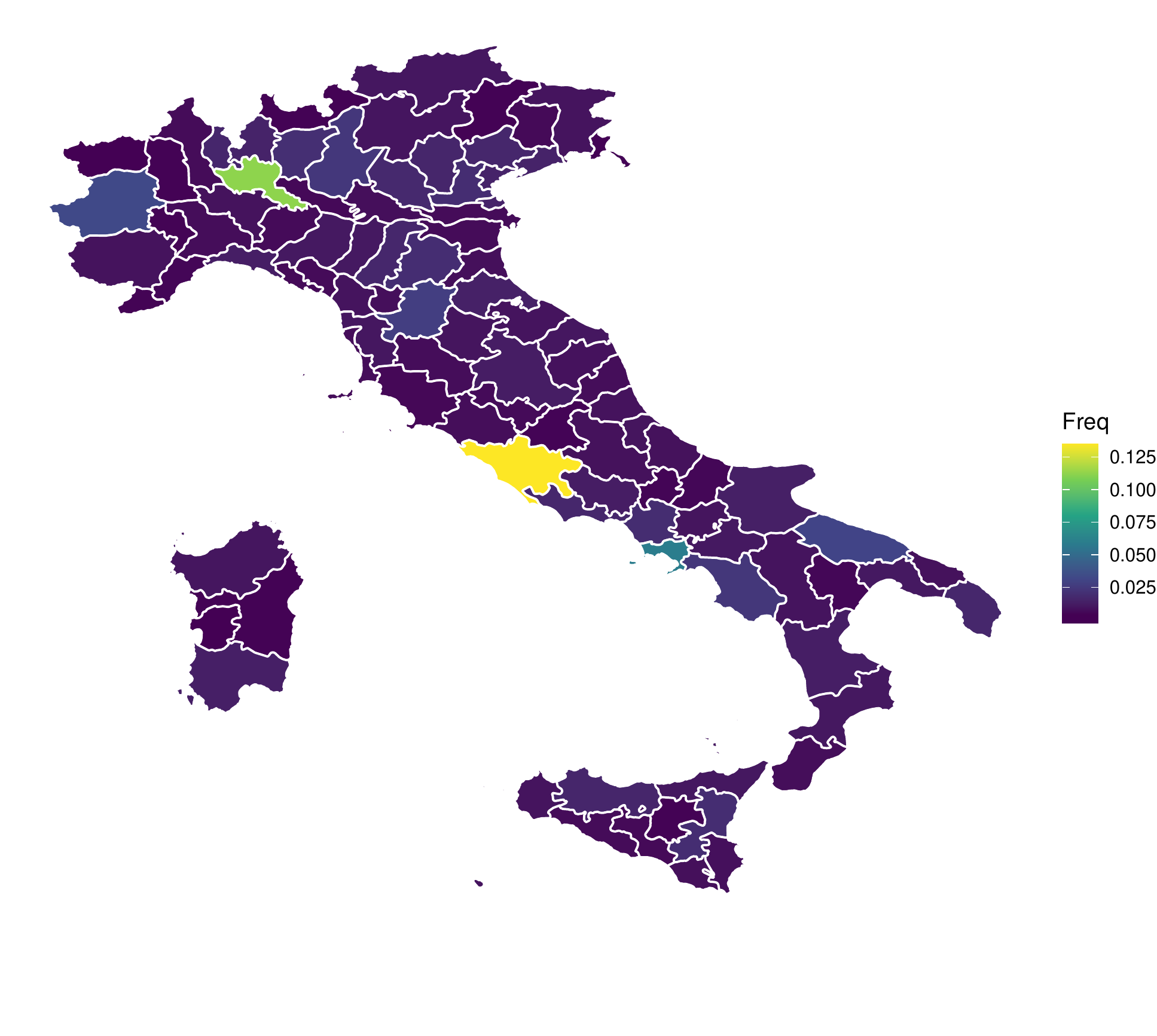}} \caption{Geo-localization of the 2013 sample. The left panel shows INNs are presented in the left panel, while NOINNs in the right one}\label{fig:1}
\end{center}
\end{figure}

Further descriptive statistics on the 2013 sample are here proposed. We study the distribution of employments (Figure~\ref{fig:0}) and EBITDA (Figure~\ref{fig:0ebi} and Table~\ref{tab:innov7}) 
in the two first years of activity (2013 and 2014) according to the inscription (or not) on the special section in the Italian companies register. We also compare the geographical distribution, at NUTS3 level, of INNs and NOINNs in 2013 (see Figure~\ref{fig:1}). 

\begin{table}[htbp]
\caption{Juridical form of the 2013 innovative start-ups ($n=1010$). Source: AIDA}\centering
\begin{footnotesize}
\begin{tabular}{lrrrrrrr}
  \hline\hline Juridical Form& Number of innovative start-ups\\ 
  \hline
  S.C.A.R.L.P.A. &  14 (1.39\%)\\ 
  S.P.A. &  13 (1.29\%)\\ 
  S.R.L. & 820 (81.19\%)\\ 
  S.R.L. a capitale ridotto &  11 (1.09\%)\\ 
  S.R.L. a socio unico &  45 (4.46\%)\\ 
  S.R.L. semplificata & 106 (10.50\%)\\ 
  Societ\`a consortile a responsabilit\`a limitata &   1 (0.1\%)\\ 
   \hline\hline
\end{tabular}
\end{footnotesize}\label{tab:innov6}
\end{table}

\begin{table}[htbp]
\caption{2013 innovative start-ups - 
EBITDA observed distribution respectively in the first (2013) and in the second (2014) year of activity. Source: AIDA}\centering
\begin{footnotesize}
\begin{tabular}{llrrrrrrrrrr}
  \hline  \hline
  Var& &Min.  &1st Qu.   &Median     &Mean  &3rd Qu.     &Max.     &NA's &t-stat&p-val \\ 
  \hline
EBITDA  &INNs &-537.40  & -3.68   &-0.39  & -3.02   & 1.95&  207.20      &213 &&\\
&&&&&&&&(21.09\%)&&\\
2013&NOINNs &-12360     &-1.53      &0.01     &11.73      &7.51 &120,000     & 9,946 &&\\
&&&&&&&&(14.78\%)&&\\
&&&&&&&&&  4.9949 &0.00\\
\hline
EBITDA  &INNs &-895.20  &-10.22    &0.09  &-11.37    &9.18  &379.40        &78 &&\\
&&&&&&&&(7.72\%)&&\\
2014&NOINNs &-18090    & -1.18     & 6.43  &   37.66     &25.43 &261,000   &  11,110 &&
\\
&&&&&&&&(16.51\%)&&\\
&&&&&&&&&  7.7044& 0.00\\\hline\hline
\end{tabular}
\end{footnotesize}\label{tab:innov7}
\end{table}

\end{document}